\documentclass[%
 reprint,
 amsmath,amssymb,
 aps,
prb,
]{revtex4-2}

\usepackage{graphicx}
\usepackage{dcolumn}
\usepackage{bm}
\usepackage{upgreek}
\usepackage{xcolor}
\usepackage{cases}
\usepackage{soul}
\usepackage{hyperref}

\begin{document}

\preprint{APS/123-QED}

\title{Inverse-Faraday effect from the orbital angular momentum of light}

\author{Vage Karakhanyan}
 \email{vage.karakhanyan.femto-st.fr}
\author{Cl\'ement Eustache}%
 \email{clement.eustache@femto-st.fr}
\author{Yannick Lefier}
 \email{yannick.lefier@gmail.com}
\author{Thierry Grosjean}
 \email{thierry.grosjean@univ-fcomte.fr}
\affiliation{%
Optics Department – FEMTO-ST Institute \\UMR 6174 - University of Bourgogne Franche-Comt\'e – CNRS - Besançon, France.
}%
\date{\today}

\begin{abstract}
It is usually admitted that the inverse Faraday effect (IFE) originates from the spin angular momentum (SAM) of light. In this paper, we evidence that part of the IFE in a metal is induced by the orbital angular momentum (OAM) of light. On the basis of a hydrodynamic model of the conduction electron gas, we describe the dependence of the IFE on the spin and orbital angular momenta as well as spin-orbit interaction  in a non-paraxial light beam.  We also numerically quantify the relative contributions of the SAM and OAM of light to the IFE in a thin gold film illuminated by different focused beams carrying SAM and/or OAM. The OAM of light provides a new degree of freedom in the control of the IFE and resulting optomagnetic field, thus potentially impacting various research fields including all-optical magnetization switching and spin-wave excitation.
\end{abstract}

\maketitle


\section{Introduction}\label{sec:Intro}

Light is known to possess polarization and spatial degrees of freedom, manifested by its linear momentum as well as spin and orbital angular momenta\cite{allen:pra92}. Remarkably, the spin angular momentum (SAM)  of light can be transferred to electrons in matter, a phenomenon which refers to as the inverse Faraday effect (IFE) \cite{pershan:pr66,popova:prb11,hertel:jmmm06}. The inverse Faraday effect (IFE) has attracted much attention for its ability to generate light-induced magnetization, thereby opening the prospect of an ultrafast magnetic data storage \cite{beaurepaire:prl96,stanciu:prl07,kirilyuk:rmp10} and a non-contact excitation of spin-waves \cite{kimel:nature05,kalashnikova:prb08,satoh:natphot12,savochkin:scirep17,matsumoto:prb20}. Plasmonic nanostructures have recently been investigated to locally enhance and control the IFE in non-magnetic metals \cite{smolyaninov:prb05,gu:josab10,cheng:np20,koshelev:prb15,hamidi:oc15,lefier:thesis,nadarajah:ox17,hurst:prb18,mondal:prb15,karakhanyan:ol21,karakhanyan:osac21} and in hybrid structures including magnetic materials
\cite{liu:nl15,dutta:ome17,chu:ome20,ignatyeva:natcomm19,im:prb19,cheng:nl20}. 

So far, the orbital counterpart of the spin-based IFE has been hardly addressed. To our knowledge, an optomagnetism induced by the OAM of light has not been reported yet. As a first evidence of the interaction of magnetism and OAM of light,  Sirenko et al showed a vortex beam dichroism in a magnetized material at THz frequencies \cite{sirenko:prl19}. The OAM of light and its combination to the optical SAM holds promise of new opportunities and strategies in tailoring optomagnetic effects \cite{maccaferri:jap20}. 

In this paper, we provide a spin and orbital angular momentum representation of the IFE in a metal. In the case of axisymmetrical optical systems (including vortex beams), we analytically show the role of the SAM and OAM of light, as well as spin-orbit interaction (SOI), in the generation of an opto-induced magnetization. We find that the SAM contributes to the IFE only in non-paraxial optical beams (carrying noticeable longitudinal light fields), whereas the OAM has a non negligible contribution even in the paraxial optical regime. Finally, we numerically  quantify the spin and orbital parts of the IFE in a thin gold film under illumination with different focused beams carrying SAM and/or OAM. We numerically confirm the importance of the SOI of light in the IFE and resulting optomagnetic field, which manifests via SAM-to-OAM or OAM-to-SAM conversions at focus.

\section{Model Description}\label{sec:ModelDescription}

In a hydrodynamic approach, the conduction electron dynamics in a metal can be described from the Euler's equation \cite{euler1757principes,scalora:pra10}:
\begin{equation}\label{eq:Euler_form3}
m_e\frac{\partial \mathbf{v}}{\partial t} + m_e(\mathbf{v} \cdot \mathbf{\nabla}) \mathbf{v} = -  \frac{m_e}{\tau} \mathbf{v} +  e \mathbf{E} + \mu_0 e \mathbf{v} \times \mathbf{H} - m_e\frac{\beta^2}{n} \nabla n.
\end{equation}
where $m_e$, $n$, $\tau$ and $\mathbf{v}$ are the effective mass, the conduction electron fluid density, the collision time and velocity of the conduction electrons, 
respectively. $\mathbf{E}$ and $\mathbf{H}$ are the applied electric and magnetic 
optical fields. The last term in Eq. 
\eqref{eq:Euler_form3} is due to the electron gas pressure, with $\beta$ 
proportional to the Fermi velocity $v_F$. $\mu_0$ and $e$ are the permeability of free space and the elementary charge, respectively.
$\mathbf{E}(\mathbf{r},t)$, $\mathbf{H}(\mathbf{r},t)$, $\mathbf{v}(\mathbf{r},t)$, $n(\mathbf{r},t)$ are time and space dependent variables. $n$ and $\mathbf{j}$ satisfy the continuity 
equation:
\begin{equation}
    \mathbf{\nabla} \cdot \mathbf{j}= -e \frac{\partial n}{\partial t},
    \label{eq:cont}
\end{equation}
where $\mathbf{j}=n e \mathbf{v}= \partial \mathbf{P}/\partial t$ is the current density 
 and $\mathbf{P}$ is the polarization density vector.

Eq. \eqref{eq:cont} can be rewritten as:
\begin{equation}\label{eq:n}
n(\mathbf{r},t)=n_0-\dfrac{1}{e} \mathbf{\nabla}\cdot\mathbf{P},
\end{equation}
where $n_0$ is the background, equilibrium charge density
in the absence of any applied fields.

Assuming that the time variation of the conduction electron fluid density is relatively small (i.e., $\dot{n}<<n$), the ratio $\dot{n}/n$ may be expanded in powers of  $1/(n_0e )$: 
\begin{equation}\label{eq:nRatio}
\begin{split}
\dfrac{\dot{n}(r,t)}{n(r,t)}=-\dfrac{1}{n_0e}\mathbf{\nabla}\cdot\dot{\mathbf{P}}\left(  1-\dfrac{1}{n_0e} \mathbf{\nabla}\cdot\mathbf{P}   \right)^{-1} \approx\\ -\dfrac{1}{n_0e}\mathbf{\nabla}\cdot\dot{\mathbf{P}}\left(  1+\dfrac{1}{n_0e} \mathbf{\nabla}\cdot\mathbf{P}   \right).
\end{split}
\end{equation}
Thus, we have:
\begin{equation}\label{eq:nRatio2}
\dfrac{\dot{n}(r,t)}{n(r,t)}
\approx -\dfrac{1}{n_0e}\mathbf{\nabla}\cdot\dot{\mathbf{P}}
-\dfrac{1}{n^2_0e^2} (\mathbf{\nabla}\cdot\mathbf{P}  ) (\mathbf{\nabla}\cdot\dot{\mathbf{P}}).
\end{equation}
By applying the above-described assumptions in Eq. \eqref{eq:Euler_form3}, we find \cite{scalora:pra10,ciraci:prb12}: 

\begin{equation}\label{eq:hydro1final}
\begin{split}
\dfrac{\partial \mathbf{j}}{\partial t} +
\frac{\mathbf{j}}{\tau} =  
\frac{e^2n_0}{m_e}\mathbf{E} -
e\beta^2 \mathbf{\nabla}n-
\frac{e}{m_e}\left( \mathbf{\nabla}\cdot\mathbf{P} \right)\mathbf{E} + \\
+\frac{\mu_0 e}{m_e} \mathbf{j} \times \mathbf{H} -
\dfrac{1}{en_0}\left[\left( \mathbf{\nabla} \cdot \mathbf{j} \right)\mathbf{j}+
(\mathbf{j} \cdot \mathbf{\nabla}) \mathbf{j}\right]
\end{split}
\end{equation}.

\section{Time harmonic regime}

In the time harmonic (i.e., monochromatic) regime,  the electric and magnetic optical fields become $\mathbf{E} = \mathbf{E_\omega}(\mathbf{r})e^{-i\omega t}+c.c$ and $\mathbf{H} = \mathbf{H_\omega}(\mathbf{r}) e^{-i\omega t}+c.c$, where $\omega$ is the angular frequency, $t$ is time, $c.c$ is the complex conjugate. To predict both the linear and nonlinear responses of the metal, we solve Eq. \eqref{eq:hydro1final} using a perturbation approach. To this end, the current density $\mathbf{j}$ is written as:

\begin{equation}
\mathbf{j}= (\mathbf{j_\omega}e^{-i\omega t} +c.c) + \mathbf{j_{NL}}.
\end{equation}

 $\mathbf{j}_{\boldsymbol{\omega}}$ and $\mathbf{j_{NL}}$ are the linear and nonlinear contributions to the current density, respectively. In the following, we focus on the nonlinear optical process defined by the drift current density $\mathbf{j_d}=\langle \mathbf{j_{NL}}\rangle$ where the operator $\langle \rangle$ denotes time averaging. $\mathbf{j_d}$ originates from an optical rectification process \cite{shen1984principles}. Because $\mathbf{j_{NL}}$ is small as compared to $\mathbf{j}_{\boldsymbol{\omega}}$, the perturbation method turns Eq \eqref{eq:hydro1final} into the following couple of equations:.

\begin{equation}
   \mathbf{j_{\omega}}  = \gamma_\omega \mathbf{E}_{\mathbf{\omega}} -
   e\beta^2 \mathbf{\nabla}n,
   \label{eq:NonlocalE}
\end{equation}

\begin{equation}
   \begin{split}
   \mathbf{j_d}  = - \frac{\tau}{n_0e} \mathbf{Re}\left[  \frac{\gamma_0 i}{\omega \tau}(\nabla \cdot \mathbf{j}_\omega)\mathbf{E^*_\omega}+\right.
    \left( \nabla \cdot \mathbf{j_\omega} \right)\mathbf{j^*_\omega}+\\
    +(\mathbf{j_\omega}\cdot \nabla) \mathbf{j^*_\omega} 
    - \mu_0 \gamma_0\mathbf{j_\omega} \times \mathbf{H_\omega^*}\left. \vphantom{\frac{\gamma_0 i}{\omega \tau}}\right],
    \end{split}
    \label{eq:Nonlocalj}
\end{equation}

where 

\begin{equation}
   \gamma_\omega = \dfrac{\gamma_0}{1-i \omega \tau}\label{subeq:CondDynam},
\end{equation}

and

\begin{equation}
   \gamma_0=\dfrac{n_0 e^2 \tau}{m_e},
\end{equation}

are the linear (i.e., dynamic) and DC conductivities, respectively. This system of equations describes both the linear and nonlinear (rectification) responses of a material to an optical field. It can be numerically solved using for instance the two-fluid plasma model solver \cite{morel2021solver}. However, these solvers require high computational power and are time consuming.

\section{Inverse Faraday effect}

Assuming that the optical response of a metal is mainly driven by its conduction electrons, the IFE can be described by the orbital magnetization defined as:    

\begin{equation}
\mathbf{M}=
\dfrac{1}{2V}\int_V\mathbf{L}dV,
\label{eq:AM_Magnetc_moment1}
\end{equation}
where:
\begin{equation}
\begin{split}
\mathbf{L} = \mathbf{r}\times\mathbf{j_d},   
\end{split}
\label{eq:AM_local1}
\end{equation}

is the opto-induced local orbital angular momentum in the conduction electron gas. 
In that case, the IFE directly depends on the opto-induced drift current density in the metal.

\subsection{IFE in the metal bulk} 

The last terms of Eqs. \eqref{eq:Euler_form3} and \eqref{eq:NonlocalE} both describe nonlocal 
effects in the metal. In the case of smooth and slowly varying charge densities $n$, it is possible to 
neglect nonlocal effects (i.e., considering $\beta \rightarrow 0$) and take into account 
only the local response \cite{ciraci:prb12}. This local response 
approximation simplifies the resolution of Eqs. \eqref{eq:NonlocalE} and \eqref{eq:Nonlocalj} 
\cite{hurst:prb18,sinha:acsphot20}. It is however only valid within the metal bulk. The extension of the local approximation method to metal surfaces will be addressed in section \ref{par:surface}.

In the framework of a local response approximation, Eqs. \eqref{eq:NonlocalE} and  \eqref{eq:Nonlocalj} become:

\begin{equation}
   \mathbf{j_{\omega}}  = \gamma_\omega \mathbf{E}_{\mathbf{\omega}},
    \label{eq:locModel_a}
\end{equation}

and

\begin{equation}
   \begin{split}
   \mathbf{j_d}  = - \frac{\tau}{n_0e} \mathbf{Re}\left[  \frac{i}{\omega \tau}(\nabla \cdot \mathbf{j}_\omega)\mathbf{j^*_\omega}
   +(\mathbf{j_\omega}\cdot \nabla) \mathbf{j^*_\omega}
   \right.\\
     - \dfrac{\mu_0 \gamma_0}{\tau}\mathbf{j_\omega} \times \mathbf{H_\omega^*} \left. \vphantom{\frac{\gamma_0 i}{\omega \tau}}\right],
    \end{split}
    \label{eq:locModel_b}
\end{equation}

respectively. Using the following vector identity:
\begin{equation}
\begin{split}
(\mathbf{j_\omega}\cdot \nabla) \mathbf{j^*_\omega}&=
\mathbf{j_\omega}\cdot \nabla \mathbf{j^*_\omega}-
\mathbf{j_\omega}\times( \nabla \times \mathbf{j^*_\omega}),\\ &=
\mathbf{j_\omega}\cdot \nabla \mathbf{j^*_\omega}+
i\omega \mu_0 \gamma^*\mathbf{j_\omega} \times \mathbf{H_\omega^*},
\end{split}
\end{equation}

Eq. \eqref{eq:locModel_b} becomes:

\begin{equation}\label{eq:locModelSimpRectif}
\begin{split}
   \mathbf{j_d}  =& - \frac{\tau}{n_0e} \vphantom{\frac{\gamma_0 i}{20}} \left(  \frac{1}{\omega \tau}\mathbf{Im}\left[\vphantom{\frac{\gamma_0 i}{20}}(\nabla \cdot \mathbf{j}_\omega)\mathbf{j^*_\omega}\right]\right.+\\&
   +\mathbf{Re}\left[\vphantom{\frac{\gamma_0 i}{20}}\mathbf{j_\omega}\cdot \nabla \mathbf{j^*_\omega}\right]
     - \dfrac{\mu_0|\gamma_\omega|^2}{\tau}\mathbf{Re}\left[\mathbf{E_\omega} \times \mathbf{H_\omega^*} \left. \vphantom{\frac{\gamma_0 i}{\omega \tau}}\right]\right).
    \end{split}
\end{equation}

When the spatial variation of $\gamma_\omega$ are small enough to be neglected, Eq. \eqref{eq:locModelSimpRectif} reads:

\begin{equation}
   \begin{split}
   \mathbf{j_d}  =& - \frac{|\gamma_\omega|^2}{n_0e} \vphantom{\frac{\gamma_0 i}{20}} \left(  \frac{1}{\omega }\mathbf{Im}\left[\vphantom{\frac{\gamma_0 i}{20}}(\nabla \cdot \mathbf{E}_\omega)\mathbf{E^*_\omega}\right]\right.+\\&
   +\tau \mathbf{Re}\left[\vphantom{\frac{\gamma_0 i}{20}}\mathbf{E_\omega}\cdot \nabla \mathbf{E^*_\omega}\right]
     - \mu_0\mathbf{Re}\left[\mathbf{E_\omega} \times \mathbf{H_\omega^*} \left. \vphantom{\frac{\gamma_0 i}{\omega \tau}}\right]\right).
     \label{eq:locModelSimp}
\end{split}
\end{equation}

Using vector identities and Maxwell's equations, Eq. \eqref{eq:locModelSimp} can be rewritten as:

\begin{equation}
\begin{split}
\mathbf{j_d} =\dfrac{ |\gamma_\omega|^2 }{n_0e \omega} \left( \vphantom{\frac{1}{2}}
-\dfrac{\tau \omega}{2} \nabla \left(|\mathbf{E_\omega}|^2 \right) +
\mathbf{Im}\left[  \mathbf{E_\omega^*}\cdot \nabla \mathbf{E_\omega} \right] 
\right.\\+\left. 
\dfrac{1}{2} 
\nabla \times \mathbf{Im} [\mathbf{E^*_\omega} \times \mathbf{E_\omega} ]   
\right)
\end{split}
\label{eq:bulkCurrent_SpinOrbit}
\end{equation}

According to Refs. \cite{berry:joa09,bekshaev2007transverse},  the last two terms of Eq. \eqref{eq:bulkCurrent_SpinOrbit} represent the orbital and spin parts of the time-averaged Poynting vector $\mathbf{\Pi}=\mathbf{Re}\left[\mathbf{E_\omega} \times \mathbf{H_\omega^*} \right] $, respectively. We have:

 \begin{equation}
 \mathbf{\Pi} = \mathbf{\Pi^{orb}} + \mathbf{\Pi^{spn}}, 
 \label{eq:P} 
 \end{equation}

where $\mathbf{\Pi^{orb}} = \mathbf{Im}\left[  \mathbf{E_\omega^*}\cdot \nabla \mathbf{E_\omega} \right]$ and 
$\mathbf{\Pi^{spn}}  = \nabla \times \mathbf{S}$.  Vector $\mathbf{S} = \mathbf{Im} [\mathbf{E_\omega^*} \times \mathbf{E_\omega}]$ is proportional to the SAM density of light \cite{bliokh:natcom14,berry:joa09,neugebauer:prl15,aiello2015transverse}. The opto-induced drift current density in the metal bulk becomes:

\begin{equation}
\mathbf{j_d} = \mathbf{j^{ig}_d}  + \mathbf{j^{orb}_d} + \mathbf{j^{spn}_d}, \label{eq:jd}
\end{equation}

where,

\begin{equation}
    \mathbf{j^{ig}_d} = -\dfrac{ \tau|\gamma_\omega|^2 }{2 n_0e} \nabla |\mathbf{E_\omega}|^2,\label{subeq:IG_term}
\end{equation} 
\begin{equation}
    \mathbf{j^{orb}_d} = \dfrac{ |\gamma_\omega|^2 }{n_0e \omega}\mathbf{\Pi^{orb}},\label{subeq:ORB_term}
\end{equation} 
\begin{equation}
    \mathbf{j^{spn}_d} = \dfrac{|\gamma_\omega|^2 }{2n_0e \omega} \mathbf{\Pi^{spn}}=
    \dfrac{|\gamma_\omega|^2 }{2n_0e \omega}\nabla \times \mathbf{S}.\label{subeq:SPN_term} 
\end{equation}

The opto-induced drift current density thus combines three source terms (Eq. \eqref{eq:jd}). Term $\mathbf{j^{ig}_d}$ is linked to the so-called intensity-gradient force
\cite{gordon1973radiation,ashkin1983stability} that is central in optical tweezing applications \cite{xu2020kerker}. The other two source terms are related to the orbital and spin parts of the Poynting vector. 

In the monochromatic regime, the linear momentum of light is proportional to the time averaged Poynting vector \cite{berry:joa09,bliokh:natcom14}. Eqs. \eqref{subeq:ORB_term} and \eqref{subeq:SPN_term} thus describe momentum transfer from light to the conduction electrons in a metal, leading to an optical drag effect \cite{goff:prb97}. $\mathbf{\Pi^{orb}}$ and $\mathbf{\Pi^{spn}}$ are thus proportional to the canonical (orbital) and spin momentum densities of light, respectively. When vector-multiplied by $\mathbf{r}$, the source terms of Eqs \eqref{subeq:ORB_term} and \eqref{subeq:SPN_term} give the orbital and spin contributions to the angular momentum $\mathbf{L}$ of the conduction electron gas, respectively \cite{berry:joa09,bekshaev2007transverse}. The local angular momentum (Eq. \eqref{eq:AM_local1}) and resulting orbital magnetization (Eq. \eqref{eq:AM_Magnetc_moment1}) of the conduction electrons then read:

\begin{equation}
\begin{split}
\mathbf{L} = \mathbf{r}\times\mathbf{j_d^{ig}} + \mathbf{r}\times\mathbf{j_d^{orb}} + \mathbf{r}\times\mathbf{j_d^{spn}},   
\end{split}
\label{eq:AM_local}
\end{equation}

and

\begin{equation}
\mathbf{M}=
\mathbf{M^{ig}}+
\mathbf{M^{orb}}+
\mathbf{M^{spn}},
\label{eq:AM_Magnetc_moment}
\end{equation}

respectively. $\mathbf{M^{ig}}$, $\mathbf{M^{orb}}$ and $\mathbf{M^{spn}}$ are the intensity-gradient, spin and orbital contributions to the IFE, respectively. We have $\mathbf{M^{id}}=1/(2V)\int_V\mathbf{r}\times\mathbf{j_d^{id}}dV$, where "$\mathbf{id}$" stands for "$\mathbf{ig}$", "$\mathbf{orb}$" and "$\mathbf{spn}$".  Therefore, the IFE does not solely originate from the SAM of light. Eq. \eqref{eq:AM_Magnetc_moment} shows that part of the IFE in the metal bulk relies on the transfer of OAM from a light beam to the conduction electron gas of a metal. It appears from Eqs. \eqref{subeq:ORB_term} and \eqref{eq:AM_Magnetc_moment} that the orbital part of the IFE relies on an axis-symmetrical optical drag effect induced and controlled by the OAM of light.

\subsection{IFE at metal surfaces}\label{par:surface}

As previously stated, the local approximation approach used to analytically solve Eqs. \eqref{eq:NonlocalE} and \eqref{eq:Nonlocalj} is only valid within the metal bulk. To overcome ambiguities of this simplified model at metal interfaces (where strong variations of the electron fluid density occur), an analytical method has been proposed to describe opto-induced surface currents \cite{sipe:prb80,ciraci:prb12,raza:prb11,toscano:ox12,karakhanyan:ol21,karakhanyan:osac21}. This method brings a non-local correction to the local response approximation at metal surfaces. 

The idea is to define a thin metal layer beneath interfaces, whose thickness matches Thomas-Fermi length ($\lambda_{TF}\simeq$ 0.1 nm for noble metals). This layer is considered to be a "surface layer", where the electron gas pressure and the spatial variations of the conductivity are not negligible. Out of this layer, in the metal bulk, the local model applies 
($\beta \rightarrow 0$). Within the interface layer, the component $j_{\omega}^T$ of 
the linear current density that is locally parallel to the surface preserves whereas the normal component 
$j_{\omega}^N$ decays to zero. This additional boundary condition on 
$j_{\omega}^N$, which is required to solve the nonlocal problem, is attributed to
a neglected electron "spill-out" at interfaces 
\cite{ciraci:prb12}.  

We define $\rho$ as the spatial coordinate normal to the surfaces so that the metal bulk is located at 
$\rho<0$ and the surface layer corresponds to $0<\rho<\lambda_{TF}$. In that case, we have in the surface layer
$j_{\omega}^T(\rho) \approx j_{\omega}^T(0^-)$ and $j_{\omega}^N=j_{\omega}^N(0^-) 
\sigma(\rho)$, where $\sigma$ is a decaying function defined by $\int_0^{\lambda_{TF}} 
\sigma'(\rho)d\rho=-1$ \cite{ciraci:prb12}. $\sigma'$ is the derivative with respect to $\rho$. The $\sigma(\rho)$ function has the following useful properties:

\begin{equation}
    \int_0^{\lambda_{TF}} \sigma(\rho)^m \sigma'(\rho)d\rho=-\dfrac{1}{m+1},
    \label{eq:SigmaProp1}
\end{equation}

\begin{equation}
    \int_0^{\lambda_{TF}} \sigma(\rho)^m d\rho=0.
    \label{eq:SigmaProp2}
\end{equation}

This non-local correction is applied by redefining the linear conductivity of the metal as a tensor. The linear current density (cf. Eq. \eqref{eq:locModel_a}) then becomes:  

\begin{equation}\label{eq:CondTensor}
  \mathbf{j_\omega}=\left[\begin{array}{ccc}
\gamma_\omega & 0             & 0 \\
0             & \gamma_\omega & 0 \\
0             & 0             & \gamma_\omega \sigma(\rho) 
\end{array}\right]
\left[\begin{array}{c}
E_{1}^T\\
E_{2}^T \\
E^N  
\end{array}\right]
\end{equation}
where $E_{1}^T$, $E_{2}^T$  and $E^N$ are the vector components of the optical electric field that are transverse and normal to the interfaces. 
Finally, the opto-induced drift current density at metal surfaces is deduced by replacing $\mathbf{j_{\omega}}$ in Eq. \eqref{eq:locModelSimpRectif} by its expression given in Eq. \eqref{eq:CondTensor}. 

\section{IFE in axisymmetrical optical systems}\label{sec:optomagnetism}

We now focus on axisymmetrical optical systems where both the metal structure and light intensity are axially symmetric. In that case, the IFE mainly involves a magnetization oriented along the symmetry axis  \cite{hurst:prb18,sinha:acsphot20,karakhanyan:ol21}. The azimuthal component $\left[j_d\right]_\xi$  of the drift current density (in the cylindrical coordinates $(r,\xi,z)$) then becomes the main contributor to the IFE (see Eqs. \eqref{eq:AM_Magnetc_moment1} and \eqref{eq:AM_local1}).

Since $\nabla_{\xi} |\mathbf{E_\omega}|^2$=0 in light beams showing axis-symmetrical intensity, $\mathbf{j_d^{ig}}$ has no contribution to the IFE in axis-symmetrical optical systems (i.e., $[j_d^{ig}]_\xi=0$). Only the orbital and spin parts of the opto-induced drift current density are now involved (cf. Eqs. \eqref{subeq:ORB_term} and \eqref{subeq:SPN_term}, respectively). The angular momentum of the conduction electron gas now reads:

\begin{equation}
\begin{split}
\mathbf{L} &= \mathbf{r}\times\mathbf{j_d^{orb}} + \mathbf{r}\times\mathbf{j_d^{spn}},\\ 
 &= \dfrac{|\gamma_\omega|^2 }{n_0e \omega} \left[ \mathbf{r}\times\mathbf{\Pi^{orb}} + \mathbf{r}\times\mathbf{\Pi^{spn}} \right].
 \label{eq:AM_local2}
\end{split}
\end{equation}

In the following, we consider the incoming light to be a Laguerre-Gauss 
vortex beam described by a polarization helicity $s$ ($-1 \leq s \leq 1$) and a topological
charge $l$ \cite{allen:book,andrews:book,bliokh2015spin}, respectively. Laguerre-Gauss beams fulfill the cylindrical symmetry condition required to cancel the contribution of the intensity-gradient source term $\mathbf{j_d^{ig}}$. 

\subsection{IFE in the metal bulk}\label{sec:optomag_bulk}

From Eq. \eqref{eq:bulkCurrent_SpinOrbit}, we find that the azimuthal component of the drift current density within the metal bulk reads:

\begin{eqnarray}
\left[j_{d}\right]_{\xi}^{bulk} & = & \kappa  
\left[
\dfrac{l+s}{r} |\mathbf{E_\omega}|^2  - 
\dfrac{2}{r}S_z\right] + \kappa
    \left[ \dfrac{\partial S_r}{  \partial z} -
    \dfrac{\partial S_\xi }{\partial r}\right] \label{eq:toto1}
\\
& = & \left[j^{orb}_{d}\right]_{\xi}^{bulk} + \left[j^{spn}_{d}\right]_{\xi}^{bulk} \label{eq:toto2},
\end{eqnarray}

where $\kappa=|\gamma_\omega|^2 / (n_0 e \omega)$.  $S_r$, $S_\xi$, $S_z$ are the components of the SAM density $\mathbf{S}$ of light in cylindrical coordinates. We have $S_r=\mathbf{Im}[E_\xi^*E_z]$, $S_\xi=\mathbf{Im}[E^*_zE_r]$ and $S_z=\mathbf{Im}[E_r^*E_\xi]$. 

The first term of Eq. \eqref{eq:toto1}, namely:

\begin{eqnarray}
    \left[j^{orb}_{d}\right]_{\xi}^{bulk} &=& \kappa  
    \left[\dfrac{l+s}{r} |\mathbf{E_\omega}|^2  - \dfrac{2}{r}S_z\right]\label{eq:bulkOrbit}\\
    &=& \dfrac{\kappa}{r} \left( 
    l |\mathbf{E_\omega}|^2+\left[s|\mathbf{E_\omega}|^2  - 2S_z\right]\right)\label{eq:bulkOrbit2}
\end{eqnarray}


refers to the azimuthal component of $\mathbf{j^{orb}_d}$ (cf. Eq. \eqref{subeq:ORB_term}). This part of the opto-induced drift current density evidences the contribution to the IFE of the OAM of the light within the metal. 

The second term of Eq. \eqref{eq:toto1}:
\begin{equation}
    \left[j^{spn}_{d}\right]_{\xi}^{bulk} = \kappa \left[ \dfrac{\partial S_r}{  \partial z} -
    \dfrac{\partial S_\xi }{\partial r}\right].\label{eq:bulkSpin}
\end{equation}

describes the spin part of the IFE. 

Therefore, in an axis-symmetrical optical problem, our model suggests that the IFE in the metal bulk relies on both the SAM and OAM of light. From Eq. \eqref{eq:bulkOrbit2}, the OAM manifests itself via the topological charge $l$ of the incoming vortex beam and as an intermediate between the SAM of light and the OAM of the electron gas, via optical SOI (see section \ref{sec:soi}).

\subsection{IFE at metal surfaces}

In the following, we consider the opto-induced drift current density at the surfaces perpendicular to (0z). We thus have $\rho = \pm z$, where $\rho$ is defined in section \eqref{par:surface}. From section \eqref{par:surface} and Eqs. \eqref{eq:bulkOrbit} and \eqref{eq:bulkSpin}, the orbital and spin parts of the azimuthal drift current density at metal surfaces read:

\begin{equation}
    \left[j_d^{orb}\right]_{\xi}^{surf} = \kappa  
    \left(
    \dfrac{l+s}{r} |\mathbf{E_\parallel}|^2 - \dfrac{2}{r}S_z(0^-)
    \right)\label{subeq:SurfOrbit},
\end{equation}

and

\begin{equation}
    \left[j_d^{spn}\right]_{\xi}^{surf} = \kappa 
    \left( \dfrac{\partial \sigma(\rho) S_r(0^-)}{  \partial \rho} -
    \dfrac{\partial \sigma(\rho)S_\xi(0^-) }{\partial r}\right),
    \label{subeq:SurfSpin}
\end{equation}

respectively. We have $|\mathbf{E_\parallel}|^2 = E_r(0^-)E^*_r(0^-) + E_\xi (0^-)E^*_\xi (0^-)$. The definition of the optical electric field within the surface layer follows that of the linear current density (see section \ref{par:surface}). We have $E_{\omega}^T(\rho) \approx E_{\omega}^T(0^-)$ and $E_{\omega}^N(\rho)=E_{\omega}^N(0^-) \sigma(\rho)$. 

We now define $i^{s}_\xi$ as the linear density of the surface current. We have:

\begin{equation}
    i^{s}_\xi = \int_0^{\lambda_{TF}} \left[j_d^{spn}\right]_{\xi}^{surf} + \left[j_d^{orb}\right]_{\xi}^{surf} d\rho
\end{equation}

Using Eqs. \eqref{eq:SigmaProp1} and \eqref{eq:SigmaProp2}, we find:

\begin{equation}
\begin{split}
    i^{s}_\xi =   \kappa 
    \left[
    \dfrac{l+s}{r} |\mathbf{E_\parallel}|^2\lambda_{TF} - 
    \dfrac{2\lambda_{TF}}{r}S_z(0^-) 
    -S_r(0^-)\right]. \label{eq:jd_surf_init}
\end{split}
\end{equation}

The terms containing $\lambda_{TF}$ being negligible,  we finally obtain:
\begin{equation}
    i^{s}_\xi \simeq    -\kappa S_r(0^-),
    \label{eq:jd_surf}
\end{equation}

where $S_r(0^-)=\mathbf{Im}[E_{\xi}^*(0^-) E_{z}(0^-)]$. At metal surfaces,  the IFE is mainly driven by the radial component of the SAM of light.

\subsection{Paraxial approximation, spin-orbit interaction}\label{sec:soi}

In the paraxial approximation, the optical field is considered to be purely transverse ($E_z=0$) and the SAM of light reduces to $S_z = s/2 |\mathbf{E_{\omega}}|^2$ ($S_r=S_{\xi}=0$) \cite{bliokh:natcom14}.  In that case, the second term (in brackets) of Eq. \eqref{eq:bulkOrbit2} vanishes, leading to:

\begin{equation}
    \left[j^{orb}_{d}\right]_{\xi}^{bulk} = \dfrac{\kappa}{r} l |\mathbf{E_\omega}|^2.
\end{equation}

The second term of Eq. \eqref{eq:bulkOrbit2} thus analytically describes the spin-dependence of the OAM of light in non-paraxial beams, by virtue of SOI (SAM-to-OAM conversion) \cite{bliokh2010angular,bekshaev2010simple}. We will numerically see in section \ref{sec:num} that SOI (OAM-to-SAM conversion) is also inherent in the spin part of the IFE (cf. Eq. \eqref{eq:bulkSpin}).  

The spin part of the IFE being dependent on $S_{\rho}$ and $S_{\xi}$ (see Eqs. \eqref{eq:bulkSpin} and \eqref{eq:jd_surf}), the contribution of the SAM of light to the IFE vanishes in the paraxial optical regime. According to our model, the IFE induced by paraxial optical beams is solely driven by the OAM of light, which manifests via the topological charge $l$. 

Such a property is consistent with the fact that in the paraxial approximation, the orbital momentum of light (proportional to $\Pi^{orb}$; at the origin of the OAM) is the only observable contribution to the optical momentum \cite{berry:joa09,bliokh:natcom14}. The orbital momentum is here proportional to the wave vector of the optical field and is solely responsible for energy transport\cite{berry:joa09}. The spin momentum (proportional to $\mathbf{\Pi^{spn}}$; at the origin of the SAM) represents a solenoidal current, which does not contribute to energy transport \cite{berry:joa09,bliokh:natcom14}. In paraxial approximation, the spin momentum has no observable contribution to the IFE in a metal. 


\section{Optomagnetism in a thin gold layer} \label{sec:num}

To estimate the relative contributions of the SAM and OAM in the IFE and resulting optomagnetic field, we simulate the optomagnetic response of a thin gold film under illumination with single focused light beams carrying SAM and/or OAM.

\subsection{Design and theory}

\begin{figure}[ht!]
\centering
\includegraphics[width=0.4\linewidth]{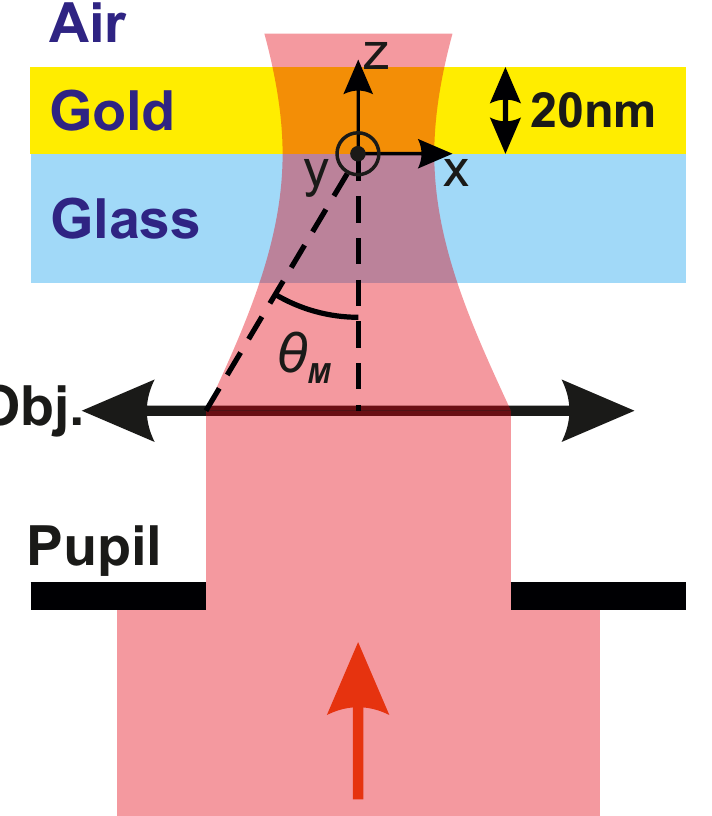}
\caption {Schematic diagram of the focusing system which involves a microscope objective of circular pupil.}
\label{fig:scheme}
\end{figure}

The configuration under study is shown in Fig. \ref{fig:scheme}. A 20-nm thick gold layer lies on a 
semi-infinite glass substrate. An incoming vortex beam of topological charge $l$, polarization helicity $s$, and of Gaussian or Laguerre-Gaussian profile is focused onto the backside of the gold film, in the substrate. The $1/e$ width of the beam waist coincides with the pupil diameter of the microscope objective. Operating in the immersion regime, the objective shows a numerical aperture $NA$ of 1.3. On the basis of the theory established by 
Richards and Wolf \cite{richards59,novotny:book}, the optical electric field at focus can be written as:

\begin{widetext}
\begin{equation}
\textbf{E}(r,\xi,z)=-\frac{i k_1 f \exp[-i k_1 f]}{2\pi} \frac{1}{\sqrt{n_1}} \times \int^{\theta_{M}}_0  \cos^{\frac{1}{2}}(\theta)\sin(\theta) F(\theta) \int^{2\pi}_0 \textbf{e}(\theta,\psi,z) \exp \left[ i \upalpha r \cos (\psi-\xi) \right] d\theta d\psi, \label{eq:sop}
\end{equation}
\end{widetext}

where $(r,\xi,z)$ are cylindrical coordinates,  $f$ is the focal length of the microscope objective, $\theta$ and $\psi$ are directional angles and $\upalpha$ is a function of $\theta$. We have $\theta_M=\arcsin(NA/n_1)$, where $n_1$ is the refractive index of the substrate. The (0z)-axis matches the symmetry axis of the microscope objective and is perpendicular to the metal surfaces (see Fig. \ref{fig:scheme}). 

Vector $\textbf{e}(\theta,\psi,z)$ takes the form: 

\begin{equation}
\textbf{e}(\theta,\psi,z)= \exp [i l \psi]
\begin{pmatrix}
E_s t_s(z) \sin \psi - E_p t^r_p(z) \cos \theta \cos \psi \\
-E_s t_s(z) \cos \psi - E_p t_p^r(z) \cos \theta \sin \psi \\
E_p t_p^z(z) \sin \theta
\end{pmatrix}
\label{eq:field1}
\end{equation}

For circular polarization ($s=\pm 1$), we have:

\begin{eqnarray}
E_s &=& -i s (\sqrt{2}/2) \exp[i s \psi],\label{eq:field2} \\ 
E_p &=& -\;(\sqrt{2}/2) \exp[i s \psi],\label{eq:field3}
\end{eqnarray}

The apodization function at the exit pupil plane of the microscope objective reads:

\begin{equation}
F(\theta)=\frac{2}{w_0}\sqrt{\frac{Z_0 P_0}{\pi |l|!}} \left( \frac{ \sqrt{2} f \sin\theta}{w_0} \right)^{|l|} \exp \left[\frac{- f^2 \sin^2\theta}{w_0^2} \right], \label{eq:apod_gauss}
\end{equation}
with :
\begin{eqnarray}
w_0 &=& f\sin \theta_M, \hspace{1,25 cm} \text{if $l$=0,} \\
w_0 &=& \frac{f \sin \theta_M}{(2|l|)^{\frac{1}{2}}}, \hspace{1.1cm} \text{  if $|l|\geq$1.}
\end{eqnarray}

$P_0$ and $w_0$ are the power and $1/e$ width of the incoming beam and $Z_0$ is the vacuum impedance. 

In the case of a radially-polarized vector vortex beams of the first order ($l=1$, $s=0$), the optical field at focus is defined by:

\begin{equation}
E_p=1 \text{, } E_s=0,
\end{equation} 

and we assume the Gaussian apodization function at the pupil plane given by Eq. \eqref{eq:apod_gauss} for $l$=0. 

For all the above-described vortex beams, the optical magnetic field is calculated by replacing $\textbf{e}(r,\xi,z)$ by $\textbf{h}(r,\xi,z)=\textbf{k}\times\textbf{e}(r,\xi,z) / \omega \mu_0$ in Eq. \eqref{eq:sop}.

Predicting the IFE in a metallic structure requires the calculation of the optical field inside the metal (see Eqs. \eqref{eq:bulkCurrent_SpinOrbit} and \eqref{eq:jd_surf}). In the basic configuration treated here, the optical field in the metal layer is described by coefficients $t_s$, $t_p^r$ and $t_p^z$ under the form $C^+ \exp[iw_2 z] + C^- \exp[-iw_2 z]$ where $w_2$ is the component of the wave vector normal to the surfaces (i.e., along (0z)). Coefficients $C^+$ and $C^-$, which are obtained by applying boundary conditions of the optical fields onto the film interfaces, can be found for instance in Ref. \cite{born:book}. The dispersion properties of gold at $\lambda$=800nm are defined by the Drude model with $\varepsilon_r = -24.76 + 0.88i$ and $\gamma_0=1.5718\cdot10^7$S.m$^{-1}$. From the calculated optical field in the metal, we anticipate the distribution of optically-induced drift current density using the formalism detailed in section \ref{sec:optomagnetism}. 

\subsection{OAM and SAM-driven optomagnetism}


In the following, we consider four different light beams carrying SAM or OAM, or a combination of SAM and OAM. We examine the circularly polarized beam ($l=0,s=1$), the radially polarized vortex beam ($l=1,s=0$) leading for instance to optical skyrmions \cite{du:np19}, and two circularly polarized vortex beams of opposite topological charges ($l=\pm 1,s=1$) \cite{zhan:ol06,zhao:prl07}. The incoming light waves are characterized by a maximum intensity of $10^{12}$W.cm$^{-2}$ at focus and a wavelength ($\lambda$) of 800 nm. The waist of the incident beam coincides with the pupil plane of the microscope objective. Its $1/e$-width matches the pupil diameter. 

\begin{figure*}[hbt!]
\centering
\includegraphics[width=0.7\linewidth]{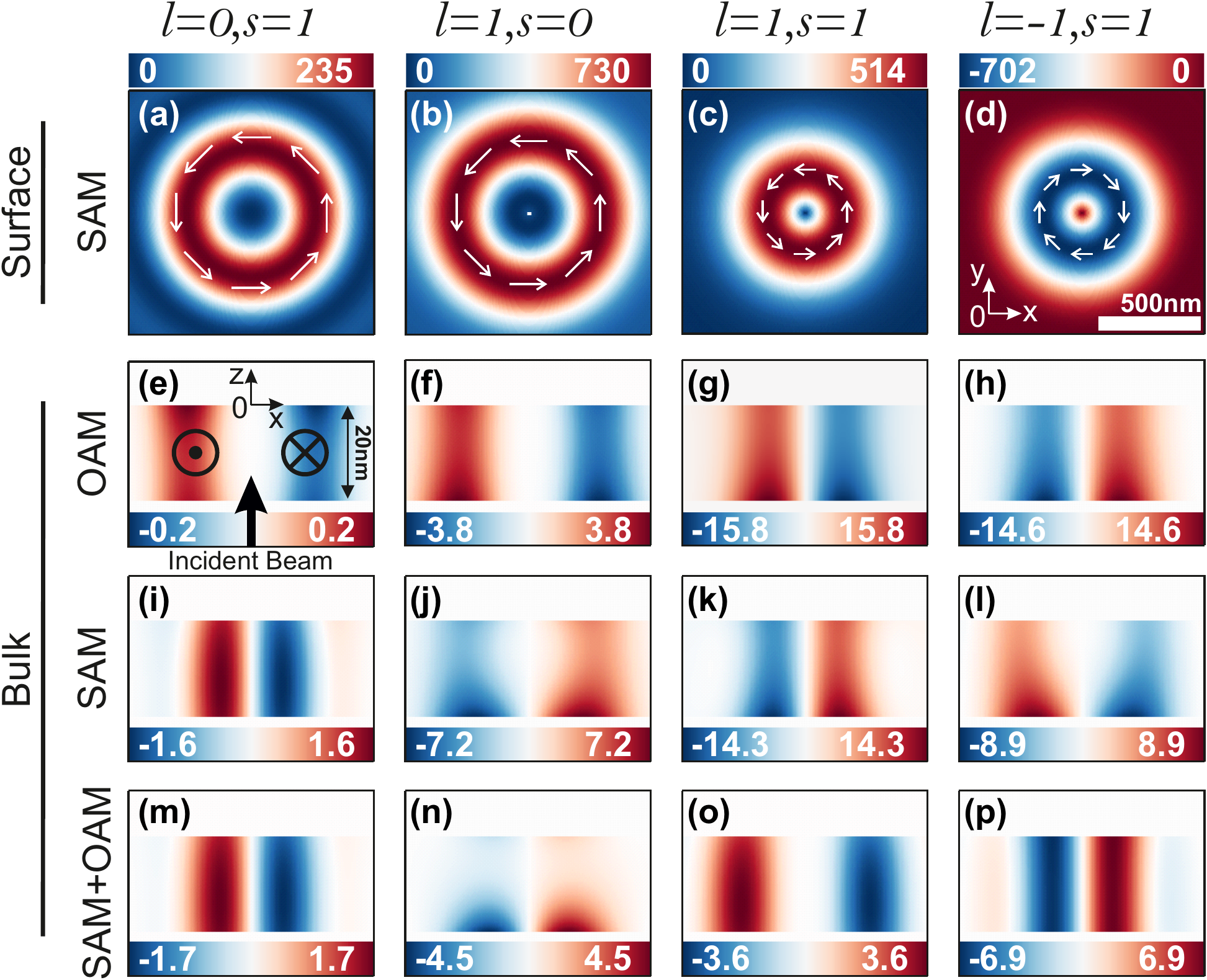}
\caption{Opto-induced current loops (a-d) at the bottom surface and (e-p) in the bulk of a thin gold film illuminated with (a,e,i,m) a circularly polarized light beam ($s$=1, $l$=0), (b,f,j,n) a radially polarized vortex beam ($s$=0, $l$=1), (c,g,k,o) a circularly polarized vortex beam ($s$=1, $l$=1), and (d,h,l,p) the same circularly polarized vortex with an opposite topological charge ($s$=1, $l$=-1). The helicity of each beam is shown with the $l$ and $s$ numbers on top of each column of the figure. The bottom surface ((xOy)-plane, see (d)) of the thin gold film matches the focal plane of the incoming light. The orientation of the surface currents in (a-d) is represented with white arrows. Eq. \eqref{subeq:SurfSpin} is used to calculate the surface currents as it is assumed to be solely driven by the SAM of light (cf. Table \ref{tab:table_magn}). Currents in the metal bulk are represented along the longitudinal (x0z)-plane (see (e)) and are decomposed in their (e-h) spin and (i-l) orbital parts (cf. Eqs. \eqref{eq:bulkSpin} and \eqref{eq:bulkOrbit}, respectively). The overall bulk current combining spin and orbital contributions is shown as well in (m-p) (cf. Eq. \eqref{eq:toto2}). The color code used to represent these out-of-plane currents is defined in (e). All currents are given in nA.}
\label{fig:Currents}
\end{figure*}

\begin{figure*}[hbt!]
\centering
\includegraphics[width=0.9\linewidth]{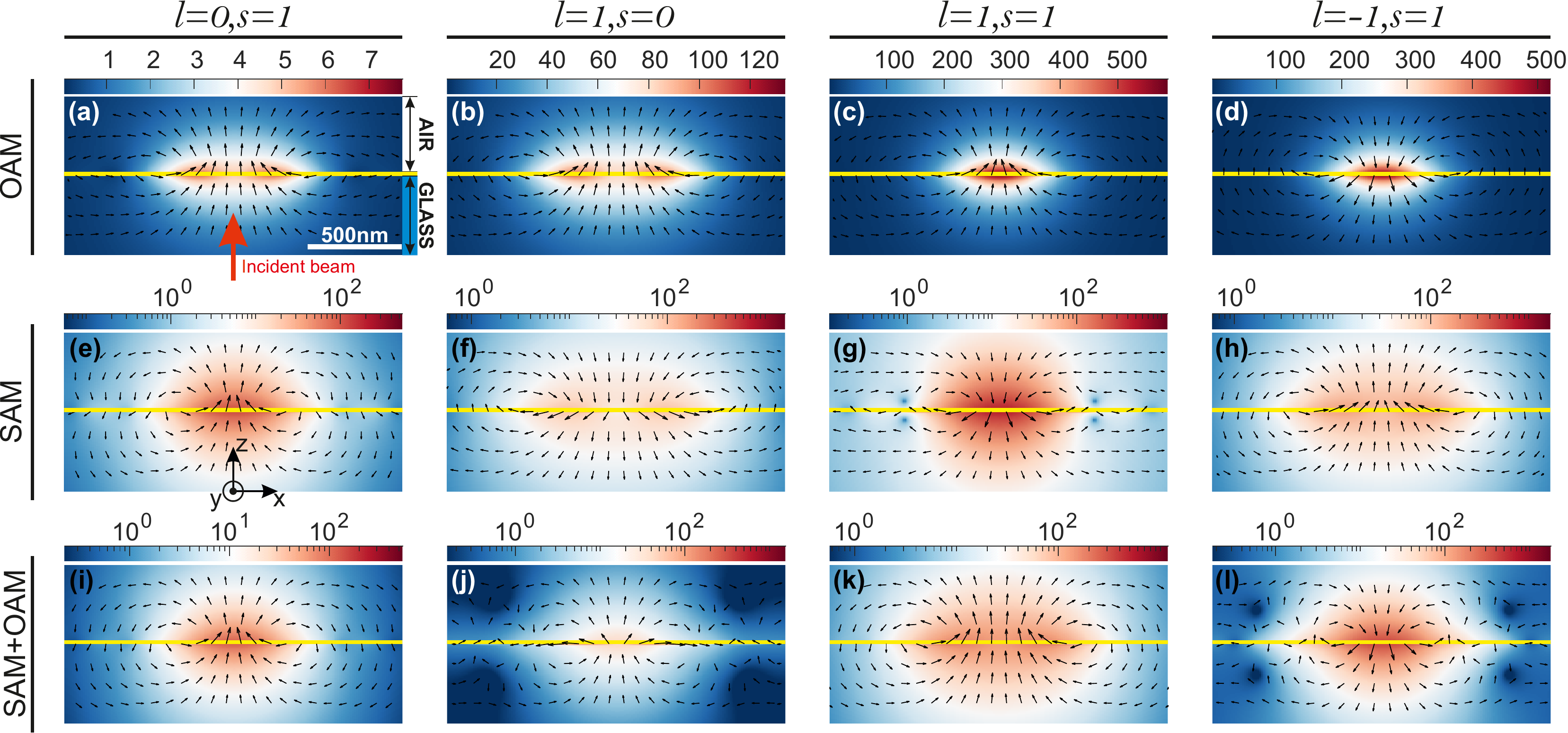}
\caption{Optically induced static magnetic field in a (x0z)-plane perpendicular to the metal surfaces. The black arrows show the local orientation of the magnetic field. (a-d) Optomagnetic field (in linear scale) from the contribution of the OAM of light to the IFE. (e-h) Optomagnetic field (in logarithmic scale) from the contribution of the SAM of light to the IFE. (i-l) Overall optomagnetic field (in logarithmic scale). In all figures, the horizontal yellow lines are used to localize the 20nm thick gold layer. The gold film is illuminated from the substrate (see indications in (a)).}
\label{fig:Optomagnetism_surfVSbulk}
\end{figure*}

Fig. \ref{fig:Currents} shows cross-sections of the bulk and surface currents generated in the metal film under illumination with the above-described light beams. The drift currents are calculated by numerically integrating the azimuthal components of the bulk and surface current densities (cf. Eqs. \eqref{eq:toto1} and \eqref{eq:jd_surf}, respectively ) over areas of 0.1$\times$0.1 nm$^2$. The optomagnetic fields originating from the SAM and OAM-driven loops of drift current are represented in Fig. \ref{fig:Optomagnetism_surfVSbulk} together with the overall optomagnetic field.

SOI \cite{bliokh:sci15} is known to tailor the helicity of focused light. As an example, in the focal region of a circularly polarized beam, the SAM is partly converted into OAM \cite{zhao:prl07}. A reciprocal OAM-to-SAM conversion occurs at the focus of vector vortex beams \cite{li:pra18,du:np19}. An immediate consequence of the SOI of light on the IFE is that opto-induced drift currents show both orbital and spin components even when the incoming beam solely carries SAM or OAM. We see from Figs. \ref{fig:Currents}(e) and \ref{fig:Currents}(i) for the circularly polarized beam (carrying SAM) and Figs. \ref{fig:Currents}(f) and \ref{fig:Currents}(j) for the radially polarized vortex beam (carrying OAM) that in the metal bulk both the orbital and spin parts of the 
drift currents are non negligible. The SAM dependence of the orbital part is analytically evidenced in Eqs. \eqref{eq:bulkOrbit} and \eqref{subeq:SurfOrbit}, whereas the OAM dependence of the spin part is contained in the spatial derivatives of Eqs. \eqref{eq:bulkSpin} and \eqref{subeq:SurfSpin}.  As anticipated in Eq. \eqref{eq:jd_surf}, the orbital part of the drift current at metal surfaces is negligible. This 
property is numerically verified in Table \ref{tab:table_magn} using Eq. \eqref{eq:jd_surf_init}. 

The effect of the SOI in the IFE is also visible in the case of circularly polarized vortex beams (carrying both SAM and OAM; cf. the two last columns of Fig. \ref{fig:Currents}). By comparing Figs. \ref{fig:Currents}(c) and \ref{fig:Currents}(d), and Figs. \ref{fig:Currents}(k) and \ref{fig:Currents}(l), we see that the spin part of the drift current in the metal bulk and at metal surfaces is reversed and shows a different morphology when the sign of the topological charge $l$ (i.e. the OAM) of the incoming beam is flipped while the polarization helicity $s$ (i.e., the SAM)  remains unchanged.


To quantify the relative contributions of the SAM and OAM of light to the IFE, we consider coefficients $\beta_{spn}$ and $\beta_{orb}$, respectively, defined as: 

\begin{equation}\label{eq:coeff_beta_spn}
    \beta_{spn} = \dfrac{\mu_{z}^{spn}}{\mu_z^{spn}+
\mu_z^{orb}},
\end{equation}

\begin{equation}\label{eq:coeff_beta_orb}
    \beta_{orb} = \dfrac{\mu_{z}^{orb}}{\mu_z^{spn}+
\mu_z^{orb}}.
\end{equation}

$\mu_{z}^{spn}$ and $\mu_{z}^{orb}$ are the $z$-component of vectors $\boldsymbol{\mu}^\mathbf{spn}$ and $\boldsymbol{\mu}^\mathbf{orb}$ defined as being derived from the spin and orbital parts of the opto-induced magnetization ($\mathbf{M^{spn}}$ and $\mathbf{M^{orb}}$, respectively).

We first focus on the ability of the SAM and OAM of light to create local orbital angular momentum in the electron gas (e-OAM), i.e., to move conduction electrons in the form of drift current loops. To this end, we define 
\begin{equation}
    \boldsymbol{\mu}^\mathbf{id}= \dfrac{1}{2V}\int_V \mathbf{r}\times|\mathbf{j_d^{id}}|dV,
    \label{eq:mu_compens}
\end{equation} 
with $|\mathbf{j_d^{id}}|=|[\mathbf{j^{id}_d}]^\mathbf{{surf}}|+|[\mathbf{j^{id}_d}]^\mathbf{bulk}|$, "$\mathbf{id}$" stands for "$\mathbf{orb}$" or "$\mathbf{spn}$". Here, $\mu_{z} = \mu_{z}^{spn} + \mu_{z}^{orb}$ quantifies the total amount of opto-induced e-OAM. We thus avoid angular momentum compensation effects that could be induced by current loops of opposite handedness to focus on how much of the total opto-induced e-OAM comes from the SAM and OAM of light. 

From these considerations, it appears that the OAM of light is the main source of e-OAM in the metal bulk for three of the four helicities of the incoming light (see Table \ref{tab:table_magn}). In contrast, the SAM of light is the main contributor to e-OAM at metal surfaces (calculated from Eq. \eqref{eq:jd_surf_init}), which confirms the simplification made in Eq. \eqref{eq:jd_surf}. Table \ref{tab:table_magn} also shows that the overall e-OAM (combining surface and bulk contributions) is mainly driven by the SAM of light. The existence of a non-negligible OAM-driven magnetization (SAM-driven magnetization, respectively) for an incoming beam solely carrying SAM (OAM, respectively) highlights the crucial role of the SOI of light in the IFE. In the case of the radially polarized vortex beam solely carrying OAM ($l=1$, $s=0$), the contribution of the SAM to the total generated e-OAM is even three times as high as the contribution of the OAM. OAM-to-SAM conversion in such a beam has already been evidenced in Ref \cite{du:np19}, leading to the concept of an optical skyrmion. When the incoming beam carries both OAM and SAM ($l=\pm 1$ and $s=1$), the OAM and SAM contributions to the overall e-OAM are almost balanced. 

\begin{table}[h]
\begin{ruledtabular}
\begin{tabular}{cccccc}
                &
                &
\textrm{l=0,s=1}&
\textrm{l=1,s=0}&
\textrm{l=1,s=1}&
\textrm{l=-1,s=1}\\
\colrule
\textbf{Bulk} &$\beta_{spn}$  & 73\% & 55\%  & 45\%   & 55\% \\
                &$\beta_{orb}$  & 27\%  & 45\%  & 55\%   & 45\% \\[4pt]
\colrule
\textbf{Surface} &$\beta_{spn}$  & 99.8\% & 99.5\%  & 99.3\%   & 99.2\% \\
                   &$\beta_{orb}$  & 0.2\%  & 0.5\%  & 0.7\%   & 0.8\% \\
                                         \colrule
\textbf{Overall} &$\beta_{spn}$  & 88\% & 76\%  &69\%   & 71\% \\
                   &$\beta_{orb}$  & 12\%  & 24\%  & 31\%   & 29\% \\
\end{tabular}
\end{ruledtabular}
\caption{$\beta_{spn}$ and $\beta_{orb}$ coefficient for four different combinations of the $l$ and $s$ parameters defining the incoming beam entering the microscope objective. $\beta_{spn}$ and $\beta_{orb}$ are defined from Eqs. \eqref{eq:coeff_beta_spn} and \eqref{eq:coeff_beta_orb}, where $\boldsymbol{\mu}^\mathbf{orb}$ and $\boldsymbol{\mu}^\mathbf{spn}$ are given in \eqref{eq:mu_compens} (total amount of e-OAM).}
\label{tab:table_magn}
\end{table}


We now study the observable magnetization associated to the IFE. $\boldsymbol{\mu}$ is then defined as the amplitude of the magnetization.:
\begin{equation}
    \boldsymbol{\mu}^\mathbf{id}=|\mathbf{M}|= \dfrac{1}{2V} \left|\int_V \mathbf{r}\times\mathbf{j_d^{id}}dV \right|,
    \label{eq:mu_observable}
\end{equation}
"$\mathbf{id}$" stands for "$\mathbf{orb}$" or "$\mathbf{spn}$".
The e-OAM compensation effects induced by current loops of opposite handedness are now taken into account. We see for instance by comparing Figs. \ref{fig:Currents} (b-d) and \ref{fig:Currents} (j-l)) that the SAM-driven drift currents at metal surfaces and bulk can be of opposite handedness. 

The SAM and OAM contributions to the overall IFE (combining surface and bulk contributions) are given in Table \ref{tab:table_magn2}. We find that except for an incoming circularly polarized beam ($l=0$, $s=1$), the contribution of the OAM of light dominates the IFE. The spin part of the IFE undergoes compensation effects due to surface and bulk currents of opposite handedness, which reduces the resulting optomagnetization. The SAM of light thus induces a larger amount of e-OAM, but due to compensation effects between surfaces and bulk contributions (electrons move in opposite directions), the SAM of light can be less efficient than its orbital counterpart to generate an observable magnetization. This surface-to-bulk compensation effect can be almost perfect, leading to a near zero contribution of the SAM to the IFE (as observed from Tables \ref{tab:table_magn} and \ref{tab:table_magn2} with the radially polarized vortex beam ($l=1$, $s=0$)).

\begin{table}[h]
\begin{ruledtabular}
\begin{tabular}{ccccc}
                &
\textrm{l=0,s=1}&
\textrm{l=1,s=0}&
\textrm{l=1,s=1}&
\textrm{l=-1,s=1}\\
\colrule
$\beta_{spn}$          & 78\%  & 0.2\%   & 36\%   & 49\% \\
$\beta_{orb}$         & 22\%  & 99.8\%   & 64\%  & 51\% \\[5pt]
\end{tabular}
\end{ruledtabular}
\caption{$\beta_{spn}$ and $\beta_{orb}$ coefficient for four different combinations of the $l$ and $s$ parameters defining the incoming beam entering the microscope objective. $\beta_{spn}$ and $\beta_{orb}$ are defined in Eqs. \eqref{eq:coeff_beta_spn} and \eqref{eq:coeff_beta_orb}, where $\boldsymbol{\mu}^\mathbf{orb}$ and $\boldsymbol{\mu}^\mathbf{spn}$ are given in \eqref{eq:mu_observable} (observable magnetization).}
\label{tab:table_magn2}
\end{table}



The resulting optomagnetic field is calculated from the drift currents shown in Fig. \ref{fig:Currents} using Biot and Savart law. We show in Fig. \ref{fig:Optomagnetism_surfVSbulk} the optomagnetic field generated from the spin and orbital parts of the drift currents (Figs. \ref{fig:Optomagnetism_surfVSbulk}(a-d) and \ref{fig:Optomagnetism_surfVSbulk}(e-h), respectively) as well as the overall magnetic field combining these two parts (Figs.\ref{fig:Optomagnetism_surfVSbulk}(i-l)). Figs. \ref{fig:Optomagnetism_surfVSbulk}(i-l) reveal that the distribution of the overall optomagnetic field strongly depends on the OAM of the incoming beam. When $l=1$ and $s=1$, the optomagnetic field forms a ring-like pattern and points upward at the center of the light beam  (Fig. \ref{fig:Optomagnetism_surfVSbulk}(k)).  The amplitude of the static magnetic field reaches 1.2 mT. When the sign of $l$ is reversed, the optomagnetic field is both flipped and confined at the beam center (Fig. \ref{fig:Optomagnetism_surfVSbulk}(i)) and it peaks at 1.8 mT.  The topological charge (i.e., OAM) of the collimated light entering the microscope objective thus provides a new degree of freedom in the control of the optomagnetism.

\section{Conclusion}

On the basis of a hydrodynamic model of the conduction electron gas, we give a spin and orbital angular momentum representation of the IFE in a metal. Both SAM and OAM contributions to the IFE in the metal bulk rely on an optical drag effect \cite{goff:prb97,noginova:prb11}: the underlying opto-induced current densities are proportional to the Poynting energy flow (optical momentum) inside the metal. In the case of an axisymmetrical optical system, we show a direct proportionality between the opto-induced drift current at metal surfaces and the radial component of the SAM of light. In the paraxial approximation, the contribution of the SAM vanishes and the IFE is solely driven by the OAM of light, which is consistent with the interpretation of the spin and orbital angular momenta of purely transverse light fields \cite{berry:joa09,bliokh:natcom14}. We also evidence that the SOI of light plays a significant role in the IFE. Finally, we numerically  quantify the relative contributions of the SAM and OAM to the IFE in a thin gold film illuminated with four different focused beams carrying SAM and/or OAM. We find that the SAM of light is the main source of drift current density regardless of the helicity of the incident light. However, compensation effects between SAM-driven surface and bulk currents of opposite handedness reduces the contribution of the SAM to the observable opto-induced magnetization. Except for circular polarization, the OAM of light is found to be the main contributor to the IFE.  We also numerically confirm the importance of the SOI of light in the IFE, which manifests both via SAM-to-OAM and OAM-to-SAM conversions at focus. The OAM of light thus opens new degrees of freedom in the control of the IFE in metals, thus potentially impacting various research fields including all-optical magnetization switching and spin-wave excitation.






\begin{thebibliography}{63}%
\makeatletter
\providecommand \@ifxundefined [1]{%
 \@ifx{#1\undefined}
}%
\providecommand \@ifnum [1]{%
 \ifnum #1\expandafter \@firstoftwo
 \else \expandafter \@secondoftwo
 \fi
}%
\providecommand \@ifx [1]{%
 \ifx #1\expandafter \@firstoftwo
 \else \expandafter \@secondoftwo
 \fi
}%
\providecommand \natexlab [1]{#1}%
\providecommand \enquote  [1]{``#1''}%
\providecommand \bibnamefont  [1]{#1}%
\providecommand \bibfnamefont [1]{#1}%
\providecommand \citenamefont [1]{#1}%
\providecommand \href@noop [0]{\@secondoftwo}%
\providecommand \href [0]{\begingroup \@sanitize@url \@href}%
\providecommand \@href[1]{\@@startlink{#1}\@@href}%
\providecommand \@@href[1]{\endgroup#1\@@endlink}%
\providecommand \@sanitize@url [0]{\catcode `\\12\catcode `\$12\catcode
  `\&12\catcode `\#12\catcode `\^12\catcode `\_12\catcode `\%12\relax}%
\providecommand \@@startlink[1]{}%
\providecommand \@@endlink[0]{}%
\providecommand \url  [0]{\begingroup\@sanitize@url \@url }%
\providecommand \@url [1]{\endgroup\@href {#1}{\urlprefix }}%
\providecommand \urlprefix  [0]{URL }%
\providecommand \Eprint [0]{\href }%
\providecommand \doibase [0]{https://doi.org/}%
\providecommand \selectlanguage [0]{\@gobble}%
\providecommand \bibinfo  [0]{\@secondoftwo}%
\providecommand \bibfield  [0]{\@secondoftwo}%
\providecommand \translation [1]{[#1]}%
\providecommand \BibitemOpen [0]{}%
\providecommand \bibitemStop [0]{}%
\providecommand \bibitemNoStop [0]{.\EOS\space}%
\providecommand \EOS [0]{\spacefactor3000\relax}%
\providecommand \BibitemShut  [1]{\csname bibitem#1\endcsname}%
\let\auto@bib@innerbib\@empty
\bibitem [{\citenamefont {Allen}\ \emph {et~al.}(1992)\citenamefont {Allen},
  \citenamefont {Beijersbergen}, \citenamefont {Spreeuw},\ and\ \citenamefont
  {Woerdman}}]{allen:pra92}%
  \BibitemOpen
  \bibfield  {author} {\bibinfo {author} {\bibfnamefont {L.}~\bibnamefont
  {Allen}}, \bibinfo {author} {\bibfnamefont {M.~W.}\ \bibnamefont
  {Beijersbergen}}, \bibinfo {author} {\bibfnamefont {R.}~\bibnamefont
  {Spreeuw}},\ and\ \bibinfo {author} {\bibfnamefont {J.}~\bibnamefont
  {Woerdman}},\ }\href@noop {} {\bibfield  {journal} {\bibinfo  {journal}
  {Phys. Rev. A}\ }\textbf {\bibinfo {volume} {45}},\ \bibinfo {pages} {8185}
  (\bibinfo {year} {1992})}\BibitemShut {NoStop}%
\bibitem [{\citenamefont {Pershan}\ \emph {et~al.}(1966)\citenamefont
  {Pershan}, \citenamefont {Van~der Ziel},\ and\ \citenamefont
  {Malmstrom}}]{pershan:pr66}%
  \BibitemOpen
  \bibfield  {author} {\bibinfo {author} {\bibfnamefont {P.}~\bibnamefont
  {Pershan}}, \bibinfo {author} {\bibfnamefont {J.}~\bibnamefont {Van~der
  Ziel}},\ and\ \bibinfo {author} {\bibfnamefont {L.}~\bibnamefont
  {Malmstrom}},\ }\href@noop {} {\bibfield  {journal} {\bibinfo  {journal}
  {Phys. Rev.}\ }\textbf {\bibinfo {volume} {143}},\ \bibinfo {pages} {574}
  (\bibinfo {year} {1966})}\BibitemShut {NoStop}%
\bibitem [{\citenamefont {Popova}\ \emph {et~al.}(2011)\citenamefont {Popova},
  \citenamefont {Bringer},\ and\ \citenamefont {Bl{\"u}gel}}]{popova:prb11}%
  \BibitemOpen
  \bibfield  {author} {\bibinfo {author} {\bibfnamefont {D.}~\bibnamefont
  {Popova}}, \bibinfo {author} {\bibfnamefont {A.}~\bibnamefont {Bringer}},\
  and\ \bibinfo {author} {\bibfnamefont {S.}~\bibnamefont {Bl{\"u}gel}},\
  }\href@noop {} {\bibfield  {journal} {\bibinfo  {journal} {Phys. Rev. B}\
  }\textbf {\bibinfo {volume} {84}},\ \bibinfo {pages} {214421} (\bibinfo
  {year} {2011})}\BibitemShut {NoStop}%
\bibitem [{\citenamefont {Hertel}(2006)}]{hertel:jmmm06}%
  \BibitemOpen
  \bibfield  {author} {\bibinfo {author} {\bibfnamefont {R.}~\bibnamefont
  {Hertel}},\ }\href@noop {} {\bibfield  {journal} {\bibinfo  {journal} {J.
  Magn. Magn. Mater.}\ }\textbf {\bibinfo {volume} {303}},\ \bibinfo {pages}
  {L1} (\bibinfo {year} {2006})}\BibitemShut {NoStop}%
\bibitem [{\citenamefont {Beaurepaire}\ \emph {et~al.}(1996)\citenamefont
  {Beaurepaire}, \citenamefont {Merle}, \citenamefont {Daunois},\ and\
  \citenamefont {Bigot}}]{beaurepaire:prl96}%
  \BibitemOpen
  \bibfield  {author} {\bibinfo {author} {\bibfnamefont {E.}~\bibnamefont
  {Beaurepaire}}, \bibinfo {author} {\bibfnamefont {J.-C.}\ \bibnamefont
  {Merle}}, \bibinfo {author} {\bibfnamefont {A.}~\bibnamefont {Daunois}},\
  and\ \bibinfo {author} {\bibfnamefont {J.-Y.}\ \bibnamefont {Bigot}},\
  }\href@noop {} {\bibfield  {journal} {\bibinfo  {journal} {Phys. Rev. Lett.}\
  }\textbf {\bibinfo {volume} {76}},\ \bibinfo {pages} {4250} (\bibinfo {year}
  {1996})}\BibitemShut {NoStop}%
\bibitem [{\citenamefont {Stanciu}\ \emph {et~al.}(2007)\citenamefont
  {Stanciu}, \citenamefont {Hansteen}, \citenamefont {Kimel}, \citenamefont
  {Kirilyuk}, \citenamefont {Tsukamoto}, \citenamefont {Itoh},\ and\
  \citenamefont {Rasing}}]{stanciu:prl07}%
  \BibitemOpen
  \bibfield  {author} {\bibinfo {author} {\bibfnamefont {C.}~\bibnamefont
  {Stanciu}}, \bibinfo {author} {\bibfnamefont {F.}~\bibnamefont {Hansteen}},
  \bibinfo {author} {\bibfnamefont {A.}~\bibnamefont {Kimel}}, \bibinfo
  {author} {\bibfnamefont {A.}~\bibnamefont {Kirilyuk}}, \bibinfo {author}
  {\bibfnamefont {A.}~\bibnamefont {Tsukamoto}}, \bibinfo {author}
  {\bibfnamefont {A.}~\bibnamefont {Itoh}},\ and\ \bibinfo {author}
  {\bibfnamefont {T.}~\bibnamefont {Rasing}},\ }\href@noop {} {\bibfield
  {journal} {\bibinfo  {journal} {Phys. Rev. Lett.}\ }\textbf {\bibinfo
  {volume} {99}},\ \bibinfo {pages} {047601} (\bibinfo {year}
  {2007})}\BibitemShut {NoStop}%
\bibitem [{\citenamefont {Kirilyuk}\ \emph {et~al.}(2010)\citenamefont
  {Kirilyuk}, \citenamefont {Kimel},\ and\ \citenamefont
  {Rasing}}]{kirilyuk:rmp10}%
  \BibitemOpen
  \bibfield  {author} {\bibinfo {author} {\bibfnamefont {A.}~\bibnamefont
  {Kirilyuk}}, \bibinfo {author} {\bibfnamefont {A.~V.}\ \bibnamefont
  {Kimel}},\ and\ \bibinfo {author} {\bibfnamefont {T.}~\bibnamefont
  {Rasing}},\ }\href@noop {} {\bibfield  {journal} {\bibinfo  {journal} {Rev.
  Mod. Phys.}\ }\textbf {\bibinfo {volume} {82}},\ \bibinfo {pages} {2731}
  (\bibinfo {year} {2010})}\BibitemShut {NoStop}%
\bibitem [{\citenamefont {Kimel}\ \emph {et~al.}(2005)\citenamefont {Kimel},
  \citenamefont {Kirilyuk}, \citenamefont {Usachev}, \citenamefont {Pisarev},
  \citenamefont {Balbashov},\ and\ \citenamefont {Rasing}}]{kimel:nature05}%
  \BibitemOpen
  \bibfield  {author} {\bibinfo {author} {\bibfnamefont {A.}~\bibnamefont
  {Kimel}}, \bibinfo {author} {\bibfnamefont {A.}~\bibnamefont {Kirilyuk}},
  \bibinfo {author} {\bibfnamefont {P.}~\bibnamefont {Usachev}}, \bibinfo
  {author} {\bibfnamefont {R.}~\bibnamefont {Pisarev}}, \bibinfo {author}
  {\bibfnamefont {A.}~\bibnamefont {Balbashov}},\ and\ \bibinfo {author}
  {\bibfnamefont {T.}~\bibnamefont {Rasing}},\ }\href@noop {} {\bibfield
  {journal} {\bibinfo  {journal} {Nature}\ }\textbf {\bibinfo {volume} {435}},\
  \bibinfo {pages} {655} (\bibinfo {year} {2005})}\BibitemShut {NoStop}%
\bibitem [{\citenamefont {Kalashnikova}\ \emph {et~al.}(2008)\citenamefont
  {Kalashnikova}, \citenamefont {Kimel}, \citenamefont {Pisarev}, \citenamefont
  {Gridnev}, \citenamefont {Usachev}, \citenamefont {Kirilyuk},\ and\
  \citenamefont {Rasing}}]{kalashnikova:prb08}%
  \BibitemOpen
  \bibfield  {author} {\bibinfo {author} {\bibfnamefont {A.}~\bibnamefont
  {Kalashnikova}}, \bibinfo {author} {\bibfnamefont {A.}~\bibnamefont {Kimel}},
  \bibinfo {author} {\bibfnamefont {R.}~\bibnamefont {Pisarev}}, \bibinfo
  {author} {\bibfnamefont {V.}~\bibnamefont {Gridnev}}, \bibinfo {author}
  {\bibfnamefont {P.}~\bibnamefont {Usachev}}, \bibinfo {author} {\bibfnamefont
  {A.}~\bibnamefont {Kirilyuk}},\ and\ \bibinfo {author} {\bibfnamefont
  {T.}~\bibnamefont {Rasing}},\ }\href@noop {} {\bibfield  {journal} {\bibinfo
  {journal} {Phys. Rev. B}\ }\textbf {\bibinfo {volume} {78}},\ \bibinfo
  {pages} {104301} (\bibinfo {year} {2008})}\BibitemShut {NoStop}%
\bibitem [{\citenamefont {Satoh}\ \emph {et~al.}(2012)\citenamefont {Satoh},
  \citenamefont {Terui}, \citenamefont {Moriya}, \citenamefont {Ivanov},
  \citenamefont {Ando}, \citenamefont {Saitoh}, \citenamefont {Shimura},\ and\
  \citenamefont {Kuroda}}]{satoh:natphot12}%
  \BibitemOpen
  \bibfield  {author} {\bibinfo {author} {\bibfnamefont {T.}~\bibnamefont
  {Satoh}}, \bibinfo {author} {\bibfnamefont {Y.}~\bibnamefont {Terui}},
  \bibinfo {author} {\bibfnamefont {R.}~\bibnamefont {Moriya}}, \bibinfo
  {author} {\bibfnamefont {B.~A.}\ \bibnamefont {Ivanov}}, \bibinfo {author}
  {\bibfnamefont {K.}~\bibnamefont {Ando}}, \bibinfo {author} {\bibfnamefont
  {E.}~\bibnamefont {Saitoh}}, \bibinfo {author} {\bibfnamefont
  {T.}~\bibnamefont {Shimura}},\ and\ \bibinfo {author} {\bibfnamefont
  {K.}~\bibnamefont {Kuroda}},\ }\href@noop {} {\bibfield  {journal} {\bibinfo
  {journal} {Nat. Photonics}\ }\textbf {\bibinfo {volume} {6}},\ \bibinfo
  {pages} {662} (\bibinfo {year} {2012})}\BibitemShut {NoStop}%
\bibitem [{\citenamefont {Savochkin}\ \emph {et~al.}(2017)\citenamefont
  {Savochkin}, \citenamefont {J{\"a}ckl}, \citenamefont {Belotelov},
  \citenamefont {Akimov}, \citenamefont {Kozhaev}, \citenamefont {Sylgacheva},
  \citenamefont {Chernov}, \citenamefont {Shaposhnikov}, \citenamefont
  {Prokopov}, \citenamefont {Berzhansky} \emph {et~al.}}]{savochkin:scirep17}%
  \BibitemOpen
  \bibfield  {author} {\bibinfo {author} {\bibfnamefont {I.}~\bibnamefont
  {Savochkin}}, \bibinfo {author} {\bibfnamefont {M.}~\bibnamefont
  {J{\"a}ckl}}, \bibinfo {author} {\bibfnamefont {V.}~\bibnamefont
  {Belotelov}}, \bibinfo {author} {\bibfnamefont {I.}~\bibnamefont {Akimov}},
  \bibinfo {author} {\bibfnamefont {M.}~\bibnamefont {Kozhaev}}, \bibinfo
  {author} {\bibfnamefont {D.}~\bibnamefont {Sylgacheva}}, \bibinfo {author}
  {\bibfnamefont {A.}~\bibnamefont {Chernov}}, \bibinfo {author} {\bibfnamefont
  {A.}~\bibnamefont {Shaposhnikov}}, \bibinfo {author} {\bibfnamefont
  {A.}~\bibnamefont {Prokopov}}, \bibinfo {author} {\bibfnamefont
  {V.}~\bibnamefont {Berzhansky}}, \emph {et~al.},\ }\href@noop {} {\bibfield
  {journal} {\bibinfo  {journal} {Sci. Rep.}\ }\textbf {\bibinfo {volume}
  {7}},\ \bibinfo {pages} {1} (\bibinfo {year} {2017})}\BibitemShut {NoStop}%
\bibitem [{\citenamefont {Matsumoto}\ \emph {et~al.}(2020)\citenamefont
  {Matsumoto}, \citenamefont {Yoshimine}, \citenamefont {Himeno}, \citenamefont
  {Shimura},\ and\ \citenamefont {Satoh}}]{matsumoto:prb20}%
  \BibitemOpen
  \bibfield  {author} {\bibinfo {author} {\bibfnamefont {K.}~\bibnamefont
  {Matsumoto}}, \bibinfo {author} {\bibfnamefont {I.}~\bibnamefont
  {Yoshimine}}, \bibinfo {author} {\bibfnamefont {K.}~\bibnamefont {Himeno}},
  \bibinfo {author} {\bibfnamefont {T.}~\bibnamefont {Shimura}},\ and\ \bibinfo
  {author} {\bibfnamefont {T.}~\bibnamefont {Satoh}},\ }\href@noop {}
  {\bibfield  {journal} {\bibinfo  {journal} {Phys. Rev. B}\ }\textbf {\bibinfo
  {volume} {101}},\ \bibinfo {pages} {184407} (\bibinfo {year}
  {2020})}\BibitemShut {NoStop}%
\bibitem [{\citenamefont {Smolyaninov}\ \emph {et~al.}(2005)\citenamefont
  {Smolyaninov}, \citenamefont {Davis}, \citenamefont {Smolyaninova},
  \citenamefont {Schaefer}, \citenamefont {Elliott},\ and\ \citenamefont
  {Zayats}}]{smolyaninov:prb05}%
  \BibitemOpen
  \bibfield  {author} {\bibinfo {author} {\bibfnamefont {I.~I.}\ \bibnamefont
  {Smolyaninov}}, \bibinfo {author} {\bibfnamefont {C.~C.}\ \bibnamefont
  {Davis}}, \bibinfo {author} {\bibfnamefont {V.~N.}\ \bibnamefont
  {Smolyaninova}}, \bibinfo {author} {\bibfnamefont {D.}~\bibnamefont
  {Schaefer}}, \bibinfo {author} {\bibfnamefont {J.}~\bibnamefont {Elliott}},\
  and\ \bibinfo {author} {\bibfnamefont {A.~V.}\ \bibnamefont {Zayats}},\
  }\href@noop {} {\bibfield  {journal} {\bibinfo  {journal} {Phys. Rev. B}\
  }\textbf {\bibinfo {volume} {71}},\ \bibinfo {pages} {035425} (\bibinfo
  {year} {2005})}\BibitemShut {NoStop}%
\bibitem [{\citenamefont {Gu}\ and\ \citenamefont {Kornev}(2010)}]{gu:josab10}%
  \BibitemOpen
  \bibfield  {author} {\bibinfo {author} {\bibfnamefont {Y.}~\bibnamefont
  {Gu}}\ and\ \bibinfo {author} {\bibfnamefont {K.~G.}\ \bibnamefont
  {Kornev}},\ }\href@noop {} {\bibfield  {journal} {\bibinfo  {journal} {JOSA
  B}\ }\textbf {\bibinfo {volume} {27}},\ \bibinfo {pages} {2165} (\bibinfo
  {year} {2010})}\BibitemShut {NoStop}%
\bibitem [{\citenamefont {Cheng}\ \emph
  {et~al.}(2020{\natexlab{a}})\citenamefont {Cheng}, \citenamefont {Son},\ and\
  \citenamefont {Sheldon}}]{cheng:np20}%
  \BibitemOpen
  \bibfield  {author} {\bibinfo {author} {\bibfnamefont {O.~H.-C.}\
  \bibnamefont {Cheng}}, \bibinfo {author} {\bibfnamefont {D.~H.}\ \bibnamefont
  {Son}},\ and\ \bibinfo {author} {\bibfnamefont {M.}~\bibnamefont {Sheldon}},\
  }\href@noop {} {\bibfield  {journal} {\bibinfo  {journal} {Nat. Photonics}\
  }\textbf {\bibinfo {volume} {14}},\ \bibinfo {pages} {365} (\bibinfo {year}
  {2020}{\natexlab{a}})}\BibitemShut {NoStop}%
\bibitem [{\citenamefont {Koshelev}\ \emph {et~al.}(2015)\citenamefont
  {Koshelev}, \citenamefont {Kachorovskii},\ and\ \citenamefont
  {Titov}}]{koshelev:prb15}%
  \BibitemOpen
  \bibfield  {author} {\bibinfo {author} {\bibfnamefont {K.}~\bibnamefont
  {Koshelev}}, \bibinfo {author} {\bibfnamefont {V.~Y.}\ \bibnamefont
  {Kachorovskii}},\ and\ \bibinfo {author} {\bibfnamefont {M.}~\bibnamefont
  {Titov}},\ }\href@noop {} {\bibfield  {journal} {\bibinfo  {journal} {Phys.
  Rev. B}\ }\textbf {\bibinfo {volume} {92}},\ \bibinfo {pages} {235426}
  (\bibinfo {year} {2015})}\BibitemShut {NoStop}%
\bibitem [{\citenamefont {Hamidi}\ \emph {et~al.}(2015)\citenamefont {Hamidi},
  \citenamefont {Razavinia},\ and\ \citenamefont {Tehranchi}}]{hamidi:oc15}%
  \BibitemOpen
  \bibfield  {author} {\bibinfo {author} {\bibfnamefont {S.}~\bibnamefont
  {Hamidi}}, \bibinfo {author} {\bibfnamefont {M.}~\bibnamefont {Razavinia}},\
  and\ \bibinfo {author} {\bibfnamefont {M.}~\bibnamefont {Tehranchi}},\
  }\href@noop {} {\bibfield  {journal} {\bibinfo  {journal} {Opt. Commun.}\
  }\textbf {\bibinfo {volume} {338}},\ \bibinfo {pages} {240} (\bibinfo {year}
  {2015})}\BibitemShut {NoStop}%
\bibitem [{\citenamefont {Lefier}(2016)}]{lefier:thesis}%
  \BibitemOpen
  \bibfield  {author} {\bibinfo {author} {\bibfnamefont {Y.}~\bibnamefont
  {Lefier}},\ }\emph {\bibinfo {title} {Etudes du couplage spin-orbite en
  nano-photonique. applications {\`a} l'excitation unidirectionnelle de modes
  plasmoniques guid{\'e}s et {\`a} la g{\'e}n{\'e}ration d'opto-aimants
  nanom{\'e}triques contr{\^o}lables par l'{\'e}tat de polarisation de la
  lumi{\`e}re}},\ \href@noop {} {Ph.D. thesis},\ \bibinfo  {school}
  {Universit{\'e} de Franche-Comt{\'e}} (\bibinfo {year} {2016})\BibitemShut
  {NoStop}%
\bibitem [{\citenamefont {Nadarajah}\ and\ \citenamefont
  {Sheldon}(2017)}]{nadarajah:ox17}%
  \BibitemOpen
  \bibfield  {author} {\bibinfo {author} {\bibfnamefont {A.}~\bibnamefont
  {Nadarajah}}\ and\ \bibinfo {author} {\bibfnamefont {M.~T.}\ \bibnamefont
  {Sheldon}},\ }\href@noop {} {\bibfield  {journal} {\bibinfo  {journal} {Opt.
  Express}\ }\textbf {\bibinfo {volume} {25}},\ \bibinfo {pages} {12753}
  (\bibinfo {year} {2017})}\BibitemShut {NoStop}%
\bibitem [{\citenamefont {Hurst}\ \emph {et~al.}(2018)\citenamefont {Hurst},
  \citenamefont {Oppeneer}, \citenamefont {Manfredi},\ and\ \citenamefont
  {Hervieux}}]{hurst:prb18}%
  \BibitemOpen
  \bibfield  {author} {\bibinfo {author} {\bibfnamefont {J.}~\bibnamefont
  {Hurst}}, \bibinfo {author} {\bibfnamefont {P.~M.}\ \bibnamefont {Oppeneer}},
  \bibinfo {author} {\bibfnamefont {G.}~\bibnamefont {Manfredi}},\ and\
  \bibinfo {author} {\bibfnamefont {P.-A.}\ \bibnamefont {Hervieux}},\
  }\href@noop {} {\bibfield  {journal} {\bibinfo  {journal} {Phys. Rev. B}\
  }\textbf {\bibinfo {volume} {98}},\ \bibinfo {pages} {134439} (\bibinfo
  {year} {2018})}\BibitemShut {NoStop}%
\bibitem [{\citenamefont {Mondal}\ \emph {et~al.}(2015)\citenamefont {Mondal},
  \citenamefont {Berritta}, \citenamefont {Paillard}, \citenamefont {Singh},
  \citenamefont {Dkhil}, \citenamefont {Oppeneer},\ and\ \citenamefont
  {Bellaiche}}]{mondal:prb15}%
  \BibitemOpen
  \bibfield  {author} {\bibinfo {author} {\bibfnamefont {R.}~\bibnamefont
  {Mondal}}, \bibinfo {author} {\bibfnamefont {M.}~\bibnamefont {Berritta}},
  \bibinfo {author} {\bibfnamefont {C.}~\bibnamefont {Paillard}}, \bibinfo
  {author} {\bibfnamefont {S.}~\bibnamefont {Singh}}, \bibinfo {author}
  {\bibfnamefont {B.}~\bibnamefont {Dkhil}}, \bibinfo {author} {\bibfnamefont
  {P.~M.}\ \bibnamefont {Oppeneer}},\ and\ \bibinfo {author} {\bibfnamefont
  {L.}~\bibnamefont {Bellaiche}},\ }\href@noop {} {\bibfield  {journal}
  {\bibinfo  {journal} {Phys. Rev. B}\ }\textbf {\bibinfo {volume} {92}},\
  \bibinfo {pages} {100402} (\bibinfo {year} {2015})}\BibitemShut {NoStop}%
\bibitem [{\citenamefont {Karakhanyan}\ \emph
  {et~al.}(2021{\natexlab{a}})\citenamefont {Karakhanyan}, \citenamefont
  {Lefier}, \citenamefont {Eustache},\ and\ \citenamefont
  {Grosjean}}]{karakhanyan:ol21}%
  \BibitemOpen
  \bibfield  {author} {\bibinfo {author} {\bibfnamefont {V.}~\bibnamefont
  {Karakhanyan}}, \bibinfo {author} {\bibfnamefont {Y.}~\bibnamefont {Lefier}},
  \bibinfo {author} {\bibfnamefont {C.}~\bibnamefont {Eustache}},\ and\
  \bibinfo {author} {\bibfnamefont {T.}~\bibnamefont {Grosjean}},\ }\href
  {https://doi.org/10.1364/OL.411108} {\bibfield  {journal} {\bibinfo
  {journal} {Opt. Lett.}\ }\textbf {\bibinfo {volume} {46}},\ \bibinfo {pages}
  {613} (\bibinfo {year} {2021}{\natexlab{a}})}\BibitemShut {NoStop}%
\bibitem [{\citenamefont {Karakhanyan}\ \emph
  {et~al.}(2021{\natexlab{b}})\citenamefont {Karakhanyan}, \citenamefont
  {Eustache}, \citenamefont {Lefier},\ and\ \citenamefont
  {Grosjean}}]{karakhanyan:osac21}%
  \BibitemOpen
  \bibfield  {author} {\bibinfo {author} {\bibfnamefont {V.}~\bibnamefont
  {Karakhanyan}}, \bibinfo {author} {\bibfnamefont {C.}~\bibnamefont
  {Eustache}}, \bibinfo {author} {\bibfnamefont {Y.}~\bibnamefont {Lefier}},\
  and\ \bibinfo {author} {\bibfnamefont {T.}~\bibnamefont {Grosjean}},\
  }\href@noop {} {\bibfield  {journal} {\bibinfo  {journal} {OSA Continuum}\
  }\textbf {\bibinfo {volume} {4}},\ \bibinfo {pages} {1598} (\bibinfo {year}
  {2021}{\natexlab{b}})}\BibitemShut {NoStop}%
\bibitem [{\citenamefont {Liu}\ \emph {et~al.}(2015)\citenamefont {Liu},
  \citenamefont {Wang}, \citenamefont {Reid}, \citenamefont {Savoini},
  \citenamefont {Wu}, \citenamefont {Koene}, \citenamefont {Granitzka},
  \citenamefont {Graves}, \citenamefont {Higley}, \citenamefont {Chen} \emph
  {et~al.}}]{liu:nl15}%
  \BibitemOpen
  \bibfield  {author} {\bibinfo {author} {\bibfnamefont {T.-M.}\ \bibnamefont
  {Liu}}, \bibinfo {author} {\bibfnamefont {T.}~\bibnamefont {Wang}}, \bibinfo
  {author} {\bibfnamefont {A.~H.}\ \bibnamefont {Reid}}, \bibinfo {author}
  {\bibfnamefont {M.}~\bibnamefont {Savoini}}, \bibinfo {author} {\bibfnamefont
  {X.}~\bibnamefont {Wu}}, \bibinfo {author} {\bibfnamefont {B.}~\bibnamefont
  {Koene}}, \bibinfo {author} {\bibfnamefont {P.}~\bibnamefont {Granitzka}},
  \bibinfo {author} {\bibfnamefont {C.~E.}\ \bibnamefont {Graves}}, \bibinfo
  {author} {\bibfnamefont {D.~J.}\ \bibnamefont {Higley}}, \bibinfo {author}
  {\bibfnamefont {Z.}~\bibnamefont {Chen}}, \emph {et~al.},\ }\href@noop {}
  {\bibfield  {journal} {\bibinfo  {journal} {Nano Lett.}\ }\textbf {\bibinfo
  {volume} {15}},\ \bibinfo {pages} {6862} (\bibinfo {year}
  {2015})}\BibitemShut {NoStop}%
\bibitem [{\citenamefont {Dutta}\ \emph {et~al.}(2017)\citenamefont {Dutta},
  \citenamefont {Kildishev}, \citenamefont {Shalaev}, \citenamefont
  {Boltasseva},\ and\ \citenamefont {Marinero}}]{dutta:ome17}%
  \BibitemOpen
  \bibfield  {author} {\bibinfo {author} {\bibfnamefont {A.}~\bibnamefont
  {Dutta}}, \bibinfo {author} {\bibfnamefont {A.~V.}\ \bibnamefont
  {Kildishev}}, \bibinfo {author} {\bibfnamefont {V.~M.}\ \bibnamefont
  {Shalaev}}, \bibinfo {author} {\bibfnamefont {A.}~\bibnamefont
  {Boltasseva}},\ and\ \bibinfo {author} {\bibfnamefont {E.~E.}\ \bibnamefont
  {Marinero}},\ }\href@noop {} {\bibfield  {journal} {\bibinfo  {journal} {Opt.
  Mat. Express}\ }\textbf {\bibinfo {volume} {7}},\ \bibinfo {pages} {4316}
  (\bibinfo {year} {2017})}\BibitemShut {NoStop}%
\bibitem [{\citenamefont {Chu}\ \emph {et~al.}(2020)\citenamefont {Chu},
  \citenamefont {Beauchamp}, \citenamefont {Shah}, \citenamefont {Boltasseva},
  \citenamefont {Shalaev}, \citenamefont {Marinero} \emph
  {et~al.}}]{chu:ome20}%
  \BibitemOpen
  \bibfield  {author} {\bibinfo {author} {\bibfnamefont {A.~H.}\ \bibnamefont
  {Chu}}, \bibinfo {author} {\bibfnamefont {B.}~\bibnamefont {Beauchamp}},
  \bibinfo {author} {\bibfnamefont {D.}~\bibnamefont {Shah}}, \bibinfo {author}
  {\bibfnamefont {A.}~\bibnamefont {Boltasseva}}, \bibinfo {author}
  {\bibfnamefont {V.~M.}\ \bibnamefont {Shalaev}}, \bibinfo {author}
  {\bibfnamefont {E.~E.}\ \bibnamefont {Marinero}}, \emph {et~al.},\
  }\href@noop {} {\bibfield  {journal} {\bibinfo  {journal} {Opt. Mat.
  Express}\ }\textbf {\bibinfo {volume} {10}},\ \bibinfo {pages} {3107}
  (\bibinfo {year} {2020})}\BibitemShut {NoStop}%
\bibitem [{\citenamefont {Ignatyeva}\ \emph {et~al.}(2019)\citenamefont
  {Ignatyeva}, \citenamefont {Davies}, \citenamefont {Sylgacheva},
  \citenamefont {Tsukamoto}, \citenamefont {Yoshikawa}, \citenamefont
  {Kapralov}, \citenamefont {Kirilyuk}, \citenamefont {Belotelov},\ and\
  \citenamefont {Kimel}}]{ignatyeva:natcomm19}%
  \BibitemOpen
  \bibfield  {author} {\bibinfo {author} {\bibfnamefont {D.}~\bibnamefont
  {Ignatyeva}}, \bibinfo {author} {\bibfnamefont {C.}~\bibnamefont {Davies}},
  \bibinfo {author} {\bibfnamefont {D.}~\bibnamefont {Sylgacheva}}, \bibinfo
  {author} {\bibfnamefont {A.}~\bibnamefont {Tsukamoto}}, \bibinfo {author}
  {\bibfnamefont {H.}~\bibnamefont {Yoshikawa}}, \bibinfo {author}
  {\bibfnamefont {P.}~\bibnamefont {Kapralov}}, \bibinfo {author}
  {\bibfnamefont {A.}~\bibnamefont {Kirilyuk}}, \bibinfo {author}
  {\bibfnamefont {V.}~\bibnamefont {Belotelov}},\ and\ \bibinfo {author}
  {\bibfnamefont {A.}~\bibnamefont {Kimel}},\ }\href@noop {} {\bibfield
  {journal} {\bibinfo  {journal} {Nat. Commun.}\ }\textbf {\bibinfo {volume}
  {10}},\ \bibinfo {pages} {1} (\bibinfo {year} {2019})}\BibitemShut {NoStop}%
\bibitem [{\citenamefont {Im}\ \emph {et~al.}(2019)\citenamefont {Im},
  \citenamefont {Pae}, \citenamefont {Ri}, \citenamefont {Ho},\ and\
  \citenamefont {Herrmann}}]{im:prb19}%
  \BibitemOpen
  \bibfield  {author} {\bibinfo {author} {\bibfnamefont {S.-J.}\ \bibnamefont
  {Im}}, \bibinfo {author} {\bibfnamefont {J.-S.}\ \bibnamefont {Pae}},
  \bibinfo {author} {\bibfnamefont {C.-S.}\ \bibnamefont {Ri}}, \bibinfo
  {author} {\bibfnamefont {K.-S.}\ \bibnamefont {Ho}},\ and\ \bibinfo {author}
  {\bibfnamefont {J.}~\bibnamefont {Herrmann}},\ }\href@noop {} {\bibfield
  {journal} {\bibinfo  {journal} {Phys. Rev. B}\ }\textbf {\bibinfo {volume}
  {99}},\ \bibinfo {pages} {041401} (\bibinfo {year} {2019})}\BibitemShut
  {NoStop}%
\bibitem [{\citenamefont {Cheng}\ \emph
  {et~al.}(2020{\natexlab{b}})\citenamefont {Cheng}, \citenamefont {Wang},
  \citenamefont {Su}, \citenamefont {Wang}, \citenamefont {Cai}, \citenamefont
  {Sun},\ and\ \citenamefont {Liu}}]{cheng:nl20}%
  \BibitemOpen
  \bibfield  {author} {\bibinfo {author} {\bibfnamefont {F.}~\bibnamefont
  {Cheng}}, \bibinfo {author} {\bibfnamefont {C.}~\bibnamefont {Wang}},
  \bibinfo {author} {\bibfnamefont {Z.}~\bibnamefont {Su}}, \bibinfo {author}
  {\bibfnamefont {X.}~\bibnamefont {Wang}}, \bibinfo {author} {\bibfnamefont
  {Z.}~\bibnamefont {Cai}}, \bibinfo {author} {\bibfnamefont {N.~X.}\
  \bibnamefont {Sun}},\ and\ \bibinfo {author} {\bibfnamefont {Y.}~\bibnamefont
  {Liu}},\ }\href@noop {} {\bibfield  {journal} {\bibinfo  {journal} {Nano
  Lett.}\ }\textbf {\bibinfo {volume} {20}},\ \bibinfo {pages} {6437} (\bibinfo
  {year} {2020}{\natexlab{b}})}\BibitemShut {NoStop}%
\bibitem [{\citenamefont {Sirenko}\ \emph {et~al.}(2019)\citenamefont
  {Sirenko}, \citenamefont {Marsik}, \citenamefont {Bernhard}, \citenamefont
  {Stanislavchuk}, \citenamefont {Kiryukhin},\ and\ \citenamefont
  {Cheong}}]{sirenko:prl19}%
  \BibitemOpen
  \bibfield  {author} {\bibinfo {author} {\bibfnamefont {A.}~\bibnamefont
  {Sirenko}}, \bibinfo {author} {\bibfnamefont {P.}~\bibnamefont {Marsik}},
  \bibinfo {author} {\bibfnamefont {C.}~\bibnamefont {Bernhard}}, \bibinfo
  {author} {\bibfnamefont {T.}~\bibnamefont {Stanislavchuk}}, \bibinfo {author}
  {\bibfnamefont {V.}~\bibnamefont {Kiryukhin}},\ and\ \bibinfo {author}
  {\bibfnamefont {S.-W.}\ \bibnamefont {Cheong}},\ }\href@noop {} {\bibfield
  {journal} {\bibinfo  {journal} {Phys. Rev. Lett.}\ }\textbf {\bibinfo
  {volume} {122}},\ \bibinfo {pages} {237401} (\bibinfo {year}
  {2019})}\BibitemShut {NoStop}%
\bibitem [{\citenamefont {Maccaferri}\ \emph {et~al.}(2020)\citenamefont
  {Maccaferri}, \citenamefont {Zubritskaya}, \citenamefont {Razdolski},
  \citenamefont {Chioar}, \citenamefont {Belotelov}, \citenamefont {Kapaklis},
  \citenamefont {Oppeneer},\ and\ \citenamefont {Dmitriev}}]{maccaferri:jap20}%
  \BibitemOpen
  \bibfield  {author} {\bibinfo {author} {\bibfnamefont {N.}~\bibnamefont
  {Maccaferri}}, \bibinfo {author} {\bibfnamefont {I.}~\bibnamefont
  {Zubritskaya}}, \bibinfo {author} {\bibfnamefont {I.}~\bibnamefont
  {Razdolski}}, \bibinfo {author} {\bibfnamefont {I.-A.}\ \bibnamefont
  {Chioar}}, \bibinfo {author} {\bibfnamefont {V.}~\bibnamefont {Belotelov}},
  \bibinfo {author} {\bibfnamefont {V.}~\bibnamefont {Kapaklis}}, \bibinfo
  {author} {\bibfnamefont {P.~M.}\ \bibnamefont {Oppeneer}},\ and\ \bibinfo
  {author} {\bibfnamefont {A.}~\bibnamefont {Dmitriev}},\ }\href@noop {}
  {\bibfield  {journal} {\bibinfo  {journal} {J. Appl. Phys.}\ }\textbf
  {\bibinfo {volume} {127}},\ \bibinfo {pages} {080903} (\bibinfo {year}
  {2020})}\BibitemShut {NoStop}%
\bibitem [{\citenamefont {Euler}(1757)}]{euler1757principes}%
  \BibitemOpen
  \bibfield  {author} {\bibinfo {author} {\bibfnamefont {L.}~\bibnamefont
  {Euler}},\ }\href@noop {} {\bibfield  {journal} {\bibinfo  {journal}
  {M{\'e}moires de l'Acad{\'e}mie des Sciences de Berlin}\ ,\ \bibinfo {pages}
  {274}} (\bibinfo {year} {1757})}\BibitemShut {NoStop}%
\bibitem [{\citenamefont {Scalora}\ \emph {et~al.}(2010)\citenamefont
  {Scalora}, \citenamefont {Vincenti}, \citenamefont {De~Ceglia}, \citenamefont
  {Roppo}, \citenamefont {Centini}, \citenamefont {Akozbek},\ and\
  \citenamefont {Bloemer}}]{scalora:pra10}%
  \BibitemOpen
  \bibfield  {author} {\bibinfo {author} {\bibfnamefont {M.}~\bibnamefont
  {Scalora}}, \bibinfo {author} {\bibfnamefont {M.}~\bibnamefont {Vincenti}},
  \bibinfo {author} {\bibfnamefont {D.}~\bibnamefont {De~Ceglia}}, \bibinfo
  {author} {\bibfnamefont {V.}~\bibnamefont {Roppo}}, \bibinfo {author}
  {\bibfnamefont {M.}~\bibnamefont {Centini}}, \bibinfo {author} {\bibfnamefont
  {N.}~\bibnamefont {Akozbek}},\ and\ \bibinfo {author} {\bibfnamefont
  {M.}~\bibnamefont {Bloemer}},\ }\href@noop {} {\bibfield  {journal} {\bibinfo
   {journal} {Phys. Rev. A}\ }\textbf {\bibinfo {volume} {82}},\ \bibinfo
  {pages} {043828} (\bibinfo {year} {2010})}\BibitemShut {NoStop}%
\bibitem [{\citenamefont {Cirac{\`\i}}\ \emph {et~al.}(2012)\citenamefont
  {Cirac{\`\i}}, \citenamefont {Poutrina}, \citenamefont {Scalora},\ and\
  \citenamefont {Smith}}]{ciraci:prb12}%
  \BibitemOpen
  \bibfield  {author} {\bibinfo {author} {\bibfnamefont {C.}~\bibnamefont
  {Cirac{\`\i}}}, \bibinfo {author} {\bibfnamefont {E.}~\bibnamefont
  {Poutrina}}, \bibinfo {author} {\bibfnamefont {M.}~\bibnamefont {Scalora}},\
  and\ \bibinfo {author} {\bibfnamefont {D.~R.}\ \bibnamefont {Smith}},\
  }\href@noop {} {\bibfield  {journal} {\bibinfo  {journal} {Phys. Rev. B}\
  }\textbf {\bibinfo {volume} {85}},\ \bibinfo {pages} {201403} (\bibinfo
  {year} {2012})}\BibitemShut {NoStop}%
\bibitem [{\citenamefont {Shen}(1984)}]{shen1984principles}%
  \BibitemOpen
  \bibfield  {author} {\bibinfo {author} {\bibfnamefont {Y.-R.}\ \bibnamefont
  {Shen}},\ }\href@noop {} {\bibfield  {journal} {\bibinfo  {journal} {New
  York}\ } (\bibinfo {year} {1984})}\BibitemShut {NoStop}%
\bibitem [{\citenamefont {Morel}\ \emph {et~al.}(2021)\citenamefont {Morel},
  \citenamefont {Giust}, \citenamefont {Ardaneh},\ and\ \citenamefont
  {Courvoisier}}]{morel2021solver}%
  \BibitemOpen
  \bibfield  {author} {\bibinfo {author} {\bibfnamefont {B.}~\bibnamefont
  {Morel}}, \bibinfo {author} {\bibfnamefont {R.}~\bibnamefont {Giust}},
  \bibinfo {author} {\bibfnamefont {K.}~\bibnamefont {Ardaneh}},\ and\ \bibinfo
  {author} {\bibfnamefont {F.}~\bibnamefont {Courvoisier}},\ }\href@noop {}
  {\bibfield  {journal} {\bibinfo  {journal} {Sci. Rep.}\ }\textbf {\bibinfo
  {volume} {11}},\ \bibinfo {pages} {1} (\bibinfo {year} {2021})}\BibitemShut
  {NoStop}%
\bibitem [{\citenamefont {Sinha-Roy}\ \emph {et~al.}(2020)\citenamefont
  {Sinha-Roy}, \citenamefont {Hurst}, \citenamefont {Manfredi},\ and\
  \citenamefont {Hervieux}}]{sinha:acsphot20}%
  \BibitemOpen
  \bibfield  {author} {\bibinfo {author} {\bibfnamefont {R.}~\bibnamefont
  {Sinha-Roy}}, \bibinfo {author} {\bibfnamefont {J.}~\bibnamefont {Hurst}},
  \bibinfo {author} {\bibfnamefont {G.}~\bibnamefont {Manfredi}},\ and\
  \bibinfo {author} {\bibfnamefont {P.-A.}\ \bibnamefont {Hervieux}},\
  }\href@noop {} {\bibfield  {journal} {\bibinfo  {journal} {ACS Photonics}\
  }\textbf {\bibinfo {volume} {7}},\ \bibinfo {pages} {2429} (\bibinfo {year}
  {2020})}\BibitemShut {NoStop}%
\bibitem [{\citenamefont {Berry}(2009)}]{berry:joa09}%
  \BibitemOpen
  \bibfield  {author} {\bibinfo {author} {\bibfnamefont {M.~V.}\ \bibnamefont
  {Berry}},\ }\href@noop {} {\bibfield  {journal} {\bibinfo  {journal} {J.Opt.
  A}\ }\textbf {\bibinfo {volume} {11}},\ \bibinfo {pages} {094001} (\bibinfo
  {year} {2009})}\BibitemShut {NoStop}%
\bibitem [{\citenamefont {Bekshaev}\ and\ \citenamefont
  {Soskin}(2007)}]{bekshaev2007transverse}%
  \BibitemOpen
  \bibfield  {author} {\bibinfo {author} {\bibfnamefont {A.~Y.}\ \bibnamefont
  {Bekshaev}}\ and\ \bibinfo {author} {\bibfnamefont {M.}~\bibnamefont
  {Soskin}},\ }\href@noop {} {\bibfield  {journal} {\bibinfo  {journal} {Opt.
  Commun.}\ }\textbf {\bibinfo {volume} {271}},\ \bibinfo {pages} {332}
  (\bibinfo {year} {2007})}\BibitemShut {NoStop}%
\bibitem [{\citenamefont {Bliokh}\ \emph {et~al.}(2014)\citenamefont {Bliokh},
  \citenamefont {Bekshaev},\ and\ \citenamefont {Nori}}]{bliokh:natcom14}%
  \BibitemOpen
  \bibfield  {author} {\bibinfo {author} {\bibfnamefont {K.~Y.}\ \bibnamefont
  {Bliokh}}, \bibinfo {author} {\bibfnamefont {A.~Y.}\ \bibnamefont
  {Bekshaev}},\ and\ \bibinfo {author} {\bibfnamefont {F.}~\bibnamefont
  {Nori}},\ }\href@noop {} {\bibfield  {journal} {\bibinfo  {journal}
  {Nat.communications}\ }\textbf {\bibinfo {volume} {5}},\ \bibinfo {pages} {1}
  (\bibinfo {year} {2014})}\BibitemShut {NoStop}%
\bibitem [{\citenamefont {Neugebauer}\ \emph {et~al.}(2015)\citenamefont
  {Neugebauer}, \citenamefont {Bauer}, \citenamefont {Aiello},\ and\
  \citenamefont {Banzer}}]{neugebauer:prl15}%
  \BibitemOpen
  \bibfield  {author} {\bibinfo {author} {\bibfnamefont {M.}~\bibnamefont
  {Neugebauer}}, \bibinfo {author} {\bibfnamefont {T.}~\bibnamefont {Bauer}},
  \bibinfo {author} {\bibfnamefont {A.}~\bibnamefont {Aiello}},\ and\ \bibinfo
  {author} {\bibfnamefont {P.}~\bibnamefont {Banzer}},\ }\href@noop {}
  {\bibfield  {journal} {\bibinfo  {journal} {Phys. Rev. Lett.}\ }\textbf
  {\bibinfo {volume} {114}},\ \bibinfo {pages} {063901} (\bibinfo {year}
  {2015})}\BibitemShut {NoStop}%
\bibitem [{\citenamefont {Aiello}\ \emph {et~al.}(2015)\citenamefont {Aiello},
  \citenamefont {Banzer}, \citenamefont {Neugebauer},\ and\ \citenamefont
  {Leuchs}}]{aiello2015transverse}%
  \BibitemOpen
  \bibfield  {author} {\bibinfo {author} {\bibfnamefont {A.}~\bibnamefont
  {Aiello}}, \bibinfo {author} {\bibfnamefont {P.}~\bibnamefont {Banzer}},
  \bibinfo {author} {\bibfnamefont {M.}~\bibnamefont {Neugebauer}},\ and\
  \bibinfo {author} {\bibfnamefont {G.}~\bibnamefont {Leuchs}},\ }\href@noop {}
  {\bibfield  {journal} {\bibinfo  {journal} {Nature Photonics}\ }\textbf
  {\bibinfo {volume} {9}},\ \bibinfo {pages} {789} (\bibinfo {year}
  {2015})}\BibitemShut {NoStop}%
\bibitem [{\citenamefont {Gordon}(1973)}]{gordon1973radiation}%
  \BibitemOpen
  \bibfield  {author} {\bibinfo {author} {\bibfnamefont {J.~P.}\ \bibnamefont
  {Gordon}},\ }\href@noop {} {\bibfield  {journal} {\bibinfo  {journal} {Phys.
  Rev. A}\ }\textbf {\bibinfo {volume} {8}},\ \bibinfo {pages} {14} (\bibinfo
  {year} {1973})}\BibitemShut {NoStop}%
\bibitem [{\citenamefont {Ashkin}\ and\ \citenamefont
  {Gordon}(1983)}]{ashkin1983stability}%
  \BibitemOpen
  \bibfield  {author} {\bibinfo {author} {\bibfnamefont {A.}~\bibnamefont
  {Ashkin}}\ and\ \bibinfo {author} {\bibfnamefont {J.~P.}\ \bibnamefont
  {Gordon}},\ }\href@noop {} {\bibfield  {journal} {\bibinfo  {journal} {Opt.
  Lett}\ }\textbf {\bibinfo {volume} {8}},\ \bibinfo {pages} {511} (\bibinfo
  {year} {1983})}\BibitemShut {NoStop}%
\bibitem [{\citenamefont {Xu}\ \emph {et~al.}(2020)\citenamefont {Xu},
  \citenamefont {Nieto-Vesperinas}, \citenamefont {Qiu}, \citenamefont {Liu},
  \citenamefont {Gao}, \citenamefont {Zhang},\ and\ \citenamefont
  {Li}}]{xu2020kerker}%
  \BibitemOpen
  \bibfield  {author} {\bibinfo {author} {\bibfnamefont {X.}~\bibnamefont
  {Xu}}, \bibinfo {author} {\bibfnamefont {M.}~\bibnamefont
  {Nieto-Vesperinas}}, \bibinfo {author} {\bibfnamefont {C.-W.}\ \bibnamefont
  {Qiu}}, \bibinfo {author} {\bibfnamefont {X.}~\bibnamefont {Liu}}, \bibinfo
  {author} {\bibfnamefont {D.}~\bibnamefont {Gao}}, \bibinfo {author}
  {\bibfnamefont {Y.}~\bibnamefont {Zhang}},\ and\ \bibinfo {author}
  {\bibfnamefont {B.}~\bibnamefont {Li}},\ }\href@noop {} {\bibfield  {journal}
  {\bibinfo  {journal} {Laser Photonics Rev}\ }\textbf {\bibinfo {volume}
  {14}},\ \bibinfo {pages} {1900265} (\bibinfo {year} {2020})}\BibitemShut
  {NoStop}%
\bibitem [{\citenamefont {Goff}\ and\ \citenamefont
  {Schaich}(1997)}]{goff:prb97}%
  \BibitemOpen
  \bibfield  {author} {\bibinfo {author} {\bibfnamefont {J.~E.}\ \bibnamefont
  {Goff}}\ and\ \bibinfo {author} {\bibfnamefont {W.}~\bibnamefont {Schaich}},\
  }\href@noop {} {\bibfield  {journal} {\bibinfo  {journal} {Phys. Rev. B}\
  }\textbf {\bibinfo {volume} {56}},\ \bibinfo {pages} {15421} (\bibinfo {year}
  {1997})}\BibitemShut {NoStop}%
\bibitem [{\citenamefont {Sipe}\ \emph {et~al.}(1980)\citenamefont {Sipe},
  \citenamefont {So}, \citenamefont {Fukui},\ and\ \citenamefont
  {Stegeman}}]{sipe:prb80}%
  \BibitemOpen
  \bibfield  {author} {\bibinfo {author} {\bibfnamefont {J.}~\bibnamefont
  {Sipe}}, \bibinfo {author} {\bibfnamefont {V.}~\bibnamefont {So}}, \bibinfo
  {author} {\bibfnamefont {M.}~\bibnamefont {Fukui}},\ and\ \bibinfo {author}
  {\bibfnamefont {G.}~\bibnamefont {Stegeman}},\ }\href@noop {} {\bibfield
  {journal} {\bibinfo  {journal} {Phys. Rev. B}\ }\textbf {\bibinfo {volume}
  {21}},\ \bibinfo {pages} {4389} (\bibinfo {year} {1980})}\BibitemShut
  {NoStop}%
\bibitem [{\citenamefont {Raza}\ \emph {et~al.}(2011)\citenamefont {Raza},
  \citenamefont {Toscano}, \citenamefont {Jauho}, \citenamefont {Wubs},\ and\
  \citenamefont {Mortensen}}]{raza:prb11}%
  \BibitemOpen
  \bibfield  {author} {\bibinfo {author} {\bibfnamefont {S.}~\bibnamefont
  {Raza}}, \bibinfo {author} {\bibfnamefont {G.}~\bibnamefont {Toscano}},
  \bibinfo {author} {\bibfnamefont {A.-P.}\ \bibnamefont {Jauho}}, \bibinfo
  {author} {\bibfnamefont {M.}~\bibnamefont {Wubs}},\ and\ \bibinfo {author}
  {\bibfnamefont {N.~A.}\ \bibnamefont {Mortensen}},\ }\href@noop {} {\bibfield
   {journal} {\bibinfo  {journal} {Phys. Rev. B}\ }\textbf {\bibinfo {volume}
  {84}},\ \bibinfo {pages} {121412} (\bibinfo {year} {2011})}\BibitemShut
  {NoStop}%
\bibitem [{\citenamefont {Toscano}\ \emph {et~al.}(2012)\citenamefont
  {Toscano}, \citenamefont {Raza}, \citenamefont {Jauho}, \citenamefont
  {Mortensen},\ and\ \citenamefont {Wubs}}]{toscano:ox12}%
  \BibitemOpen
  \bibfield  {author} {\bibinfo {author} {\bibfnamefont {G.}~\bibnamefont
  {Toscano}}, \bibinfo {author} {\bibfnamefont {S.}~\bibnamefont {Raza}},
  \bibinfo {author} {\bibfnamefont {A.-P.}\ \bibnamefont {Jauho}}, \bibinfo
  {author} {\bibfnamefont {N.~A.}\ \bibnamefont {Mortensen}},\ and\ \bibinfo
  {author} {\bibfnamefont {M.}~\bibnamefont {Wubs}},\ }\href@noop {} {\bibfield
   {journal} {\bibinfo  {journal} {Opt. Express}\ }\textbf {\bibinfo {volume}
  {20}},\ \bibinfo {pages} {4176} (\bibinfo {year} {2012})}\BibitemShut
  {NoStop}%
\bibitem [{\citenamefont {Allen}\ \emph {et~al.}(2016)\citenamefont {Allen},
  \citenamefont {Barnett},\ and\ \citenamefont {Padgett}}]{allen:book}%
  \BibitemOpen
  \bibfield  {author} {\bibinfo {author} {\bibfnamefont {L.}~\bibnamefont
  {Allen}}, \bibinfo {author} {\bibfnamefont {S.~M.}\ \bibnamefont {Barnett}},\
  and\ \bibinfo {author} {\bibfnamefont {M.~J.}\ \bibnamefont {Padgett}},\
  }\href@noop {} {\emph {\bibinfo {title} {Optical angular momentum}}}\
  (\bibinfo  {publisher} {CRC press},\ \bibinfo {year} {2016})\BibitemShut
  {NoStop}%
\bibitem [{\citenamefont {Andrews}\ and\ \citenamefont
  {Babiker}(2012)}]{andrews:book}%
  \BibitemOpen
  \bibfield  {author} {\bibinfo {author} {\bibfnamefont {D.~L.}\ \bibnamefont
  {Andrews}}\ and\ \bibinfo {author} {\bibfnamefont {M.}~\bibnamefont
  {Babiker}},\ }\href@noop {} {\emph {\bibinfo {title} {The angular momentum of
  light}}}\ (\bibinfo  {publisher} {Cambridge University Press},\ \bibinfo
  {year} {2012})\BibitemShut {NoStop}%
\bibitem [{\citenamefont {Bliokh}\ \emph
  {et~al.}(2015{\natexlab{a}})\citenamefont {Bliokh}, \citenamefont
  {Rodr{\'\i}guez-Fortu{\~n}o}, \citenamefont {Nori},\ and\ \citenamefont
  {Zayats}}]{bliokh2015spin}%
  \BibitemOpen
  \bibfield  {author} {\bibinfo {author} {\bibfnamefont {K.~Y.}\ \bibnamefont
  {Bliokh}}, \bibinfo {author} {\bibfnamefont {F.~J.}\ \bibnamefont
  {Rodr{\'\i}guez-Fortu{\~n}o}}, \bibinfo {author} {\bibfnamefont
  {F.}~\bibnamefont {Nori}},\ and\ \bibinfo {author} {\bibfnamefont {A.~V.}\
  \bibnamefont {Zayats}},\ }\href@noop {} {\bibfield  {journal} {\bibinfo
  {journal} {Nat. Photonics}\ }\textbf {\bibinfo {volume} {9}},\ \bibinfo
  {pages} {796} (\bibinfo {year} {2015}{\natexlab{a}})}\BibitemShut {NoStop}%
\bibitem [{\citenamefont {Bliokh}\ \emph {et~al.}(2010)\citenamefont {Bliokh},
  \citenamefont {Alonso}, \citenamefont {Ostrovskaya},\ and\ \citenamefont
  {Aiello}}]{bliokh2010angular}%
  \BibitemOpen
  \bibfield  {author} {\bibinfo {author} {\bibfnamefont {K.~Y.}\ \bibnamefont
  {Bliokh}}, \bibinfo {author} {\bibfnamefont {M.~A.}\ \bibnamefont {Alonso}},
  \bibinfo {author} {\bibfnamefont {E.~A.}\ \bibnamefont {Ostrovskaya}},\ and\
  \bibinfo {author} {\bibfnamefont {A.}~\bibnamefont {Aiello}},\ }\href@noop {}
  {\bibfield  {journal} {\bibinfo  {journal} {Phys. Rev. A}\ }\textbf {\bibinfo
  {volume} {82}},\ \bibinfo {pages} {063825} (\bibinfo {year}
  {2010})}\BibitemShut {NoStop}%
\bibitem [{\citenamefont {Bekshaev}(2010)}]{bekshaev2010simple}%
  \BibitemOpen
  \bibfield  {author} {\bibinfo {author} {\bibfnamefont {A.~Y.}\ \bibnamefont
  {Bekshaev}},\ }\href@noop {} {\bibfield  {journal} {\bibinfo  {journal}
  {Cent. Eur. J. Phys.}\ }\textbf {\bibinfo {volume} {8}},\ \bibinfo {pages}
  {947} (\bibinfo {year} {2010})}\BibitemShut {NoStop}%
\bibitem [{\citenamefont {Richards}\ and\ \citenamefont
  {Wolf}(1959)}]{richards59}%
  \BibitemOpen
  \bibfield  {author} {\bibinfo {author} {\bibfnamefont {B.}~\bibnamefont
  {Richards}}\ and\ \bibinfo {author} {\bibfnamefont {E.}~\bibnamefont
  {Wolf}},\ }\href@noop {} {\bibfield  {journal} {\bibinfo  {journal}
  {Proceedings of the Royal Society of London. Series A. Mathematical and
  Physical Sciences}\ }\textbf {\bibinfo {volume} {253}},\ \bibinfo {pages}
  {358} (\bibinfo {year} {1959})}\BibitemShut {NoStop}%
\bibitem [{\citenamefont {Novotny}\ and\ \citenamefont
  {Hecht}(2012)}]{novotny:book}%
  \BibitemOpen
  \bibfield  {author} {\bibinfo {author} {\bibfnamefont {L.}~\bibnamefont
  {Novotny}}\ and\ \bibinfo {author} {\bibfnamefont {B.}~\bibnamefont
  {Hecht}},\ }\href@noop {} {\emph {\bibinfo {title} {Principles of
  nano-optics}}}\ (\bibinfo  {publisher} {Cambridge university press},\
  \bibinfo {year} {2012})\BibitemShut {NoStop}%
\bibitem [{\citenamefont {Born}\ and\ \citenamefont {Wolf}(2013)}]{born:book}%
  \BibitemOpen
  \bibfield  {author} {\bibinfo {author} {\bibfnamefont {M.}~\bibnamefont
  {Born}}\ and\ \bibinfo {author} {\bibfnamefont {E.}~\bibnamefont {Wolf}},\
  }\href@noop {} {\emph {\bibinfo {title} {Principles of optics:
  electromagnetic theory of propagation, interference and diffraction of
  light}}}\ (\bibinfo  {publisher} {Elsevier},\ \bibinfo {year}
  {2013})\BibitemShut {NoStop}%
\bibitem [{\citenamefont {Du}\ \emph {et~al.}(2019)\citenamefont {Du},
  \citenamefont {Yang}, \citenamefont {Zayats},\ and\ \citenamefont
  {Yuan}}]{du:np19}%
  \BibitemOpen
  \bibfield  {author} {\bibinfo {author} {\bibfnamefont {L.}~\bibnamefont
  {Du}}, \bibinfo {author} {\bibfnamefont {A.}~\bibnamefont {Yang}}, \bibinfo
  {author} {\bibfnamefont {A.~V.}\ \bibnamefont {Zayats}},\ and\ \bibinfo
  {author} {\bibfnamefont {X.}~\bibnamefont {Yuan}},\ }\href@noop {} {\bibfield
   {journal} {\bibinfo  {journal} {Nat. Phys.}\ }\textbf {\bibinfo {volume}
  {15}},\ \bibinfo {pages} {650} (\bibinfo {year} {2019})}\BibitemShut
  {NoStop}%
\bibitem [{\citenamefont {Zhan}(2006)}]{zhan:ol06}%
  \BibitemOpen
  \bibfield  {author} {\bibinfo {author} {\bibfnamefont {Q.}~\bibnamefont
  {Zhan}},\ }\href@noop {} {\bibfield  {journal} {\bibinfo  {journal} {Opt.
  Lett}\ }\textbf {\bibinfo {volume} {31}},\ \bibinfo {pages} {867} (\bibinfo
  {year} {2006})}\BibitemShut {NoStop}%
\bibitem [{\citenamefont {Zhao}\ \emph {et~al.}(2007)\citenamefont {Zhao},
  \citenamefont {Edgar}, \citenamefont {Jeffries}, \citenamefont {McGloin},\
  and\ \citenamefont {Chiu}}]{zhao:prl07}%
  \BibitemOpen
  \bibfield  {author} {\bibinfo {author} {\bibfnamefont {Y.}~\bibnamefont
  {Zhao}}, \bibinfo {author} {\bibfnamefont {J.~S.}\ \bibnamefont {Edgar}},
  \bibinfo {author} {\bibfnamefont {G.~D.}\ \bibnamefont {Jeffries}}, \bibinfo
  {author} {\bibfnamefont {D.}~\bibnamefont {McGloin}},\ and\ \bibinfo {author}
  {\bibfnamefont {D.~T.}\ \bibnamefont {Chiu}},\ }\href@noop {} {\bibfield
  {journal} {\bibinfo  {journal} {Phys. Rev. Lett.}\ }\textbf {\bibinfo
  {volume} {99}},\ \bibinfo {pages} {073901} (\bibinfo {year}
  {2007})}\BibitemShut {NoStop}%
\bibitem [{\citenamefont {Bliokh}\ \emph
  {et~al.}(2015{\natexlab{b}})\citenamefont {Bliokh}, \citenamefont
  {Smirnova},\ and\ \citenamefont {Nori}}]{bliokh:sci15}%
  \BibitemOpen
  \bibfield  {author} {\bibinfo {author} {\bibfnamefont {K.~Y.}\ \bibnamefont
  {Bliokh}}, \bibinfo {author} {\bibfnamefont {D.}~\bibnamefont {Smirnova}},\
  and\ \bibinfo {author} {\bibfnamefont {F.}~\bibnamefont {Nori}},\ }\href@noop
  {} {\bibfield  {journal} {\bibinfo  {journal} {Science}\ }\textbf {\bibinfo
  {volume} {348}},\ \bibinfo {pages} {1448} (\bibinfo {year}
  {2015}{\natexlab{b}})}\BibitemShut {NoStop}%
\bibitem [{\citenamefont {Li}\ \emph {et~al.}(2018)\citenamefont {Li},
  \citenamefont {Cai}, \citenamefont {Yan}, \citenamefont {Liang},
  \citenamefont {Zhang},\ and\ \citenamefont {Yao}}]{li:pra18}%
  \BibitemOpen
  \bibfield  {author} {\bibinfo {author} {\bibfnamefont {M.}~\bibnamefont
  {Li}}, \bibinfo {author} {\bibfnamefont {Y.}~\bibnamefont {Cai}}, \bibinfo
  {author} {\bibfnamefont {S.}~\bibnamefont {Yan}}, \bibinfo {author}
  {\bibfnamefont {Y.}~\bibnamefont {Liang}}, \bibinfo {author} {\bibfnamefont
  {P.}~\bibnamefont {Zhang}},\ and\ \bibinfo {author} {\bibfnamefont
  {B.}~\bibnamefont {Yao}},\ }\href@noop {} {\bibfield  {journal} {\bibinfo
  {journal} {Phys. Rev. A}\ }\textbf {\bibinfo {volume} {97}},\ \bibinfo
  {pages} {053842} (\bibinfo {year} {2018})}\BibitemShut {NoStop}%
\bibitem [{\citenamefont {Noginova}\ \emph {et~al.}(2011)\citenamefont
  {Noginova}, \citenamefont {Yakim}, \citenamefont {Soimo}, \citenamefont
  {Gu},\ and\ \citenamefont {Noginov}}]{noginova:prb11}%
  \BibitemOpen
  \bibfield  {author} {\bibinfo {author} {\bibfnamefont {N.}~\bibnamefont
  {Noginova}}, \bibinfo {author} {\bibfnamefont {A.}~\bibnamefont {Yakim}},
  \bibinfo {author} {\bibfnamefont {J.}~\bibnamefont {Soimo}}, \bibinfo
  {author} {\bibfnamefont {L.}~\bibnamefont {Gu}},\ and\ \bibinfo {author}
  {\bibfnamefont {M.}~\bibnamefont {Noginov}},\ }\href@noop {} {\bibfield
  {journal} {\bibinfo  {journal} {Phys. Rev. B}\ }\textbf {\bibinfo {volume}
  {84}},\ \bibinfo {pages} {035447} (\bibinfo {year} {2011})}\BibitemShut
  {NoStop}%
\end{thebibliography}
%

\end{document}